\documentclass[aps,prl,twocolumn,balance,superscriptaddress,floats]{revtex4}

\pdfoutput=1
\usepackage[a4paper, total={7.2in, 10in}]{geometry}
\usepackage{latexsym}
\usepackage{dcolumn}
\usepackage{amsmath}
\usepackage{epsf}
\usepackage{float}
\usepackage{hyperref}

\usepackage[pdftex]{graphicx}
\usepackage{epstopdf}
\usepackage{pdfpages}

\usepackage{upgreek}

\usepackage{xcolor}
\usepackage{soul}

\begin{document}

%%\preprint{APS/123-QED}

\title{A silicon Brillouin laser}% Force line breaks with \\

\author{Nils T. Otterstrom}
\affiliation{Department of Applied Physics, Yale University, New Haven, CT 06520 USA.}
\author{Ryan O. Behunin}
\affiliation{Department of Applied Physics, Yale University, New Haven, CT 06520 USA.}
\affiliation{Department of Physics and Astronomy, Northern Arizona University, Flagstaff, AZ 86001 USA.}
\author{Eric A. Kittlaus}
\affiliation{Department of Applied Physics, Yale University, New Haven, CT 06520 USA.}
\author{Zheng Wang}
\affiliation{Department of Electrical and Computer Engineering, University of Texas at Austin, Austin,
TX 78758 USA.}
\author{Peter T. Rakich}
\affiliation{Department of Applied Physics, Yale University, New Haven, CT 06520 USA.}
%\makeatother

\date{\today}

\begin{abstract}

Brillouin laser oscillators offer powerful and flexible dynamics as the basis for mode-locked lasers, microwave oscillators, and optical gyroscopes in a variety of optical systems. However, Brillouin interactions are exceedingly weak in conventional silicon photonic waveguides, stifling progress towards silicon-based Brillouin lasers. The recent advent of hybrid photonic-phononic waveguides has revealed Brillouin interactions to be one of the strongest and most tailorable nonlinearities in silicon. Here, we harness these engineered nonlinearities to demonstrate Brillouin lasing in silicon. Moreover, we show that this silicon-based Brillouin laser enters an intriguing regime of dynamics, in which optical self-oscillation produces phonon linewidth narrowing.  Our results provide a platform to develop a range of applications for monolithic integration within silicon photonic circuits.

\end{abstract}

\maketitle

\section{Introduction}
With the ability to control the optical and electronic properties of silicon, the field of silicon photonics has produced a variety of chip-scale optical devices \cite{Reed2010,liang2010recent,leuthold2010nonlinear} for applications ranging from high-bandwidth communications \cite{alduino2007interconnects} to biosensing on a chip \cite{devos07}. The rapid proliferation of these technologies has spurred interest in strategies to reshape the spectral and coherence properties of light for a wide array of on-chip functionalities. One promising approach to customize on-chip light involves using the nonlinear optical properties of silicon to create optical laser oscillators \cite{leuthold2010nonlinear}. For example, Raman nonlinearities have been harnessed to create all-silicon Raman lasers \cite{boyraz2004demonstration,rong2005continuous}.  Brillouin interactions, produced by the coupling between light and sound, could offer a complementary set of behaviors and capabilities for laser technologies in silicon.
By exploiting these nonlinearities in a variety of physical systems, Brillouin lasers have been designed to yield everything from frequency-tunable laser emission \cite{takuma1964stimulated} and mode-locked pulsed lasers \cite{lecoeuche1998brillouin} to low-noise oscillators and optical gyroscopes \cite{Smith91,Loh15,kabakova2013narrow,li2017microresonator}. 

\begin{figure*}[!tb]
\centering%\vspace{-10pt}
\includegraphics[width=\linewidth]{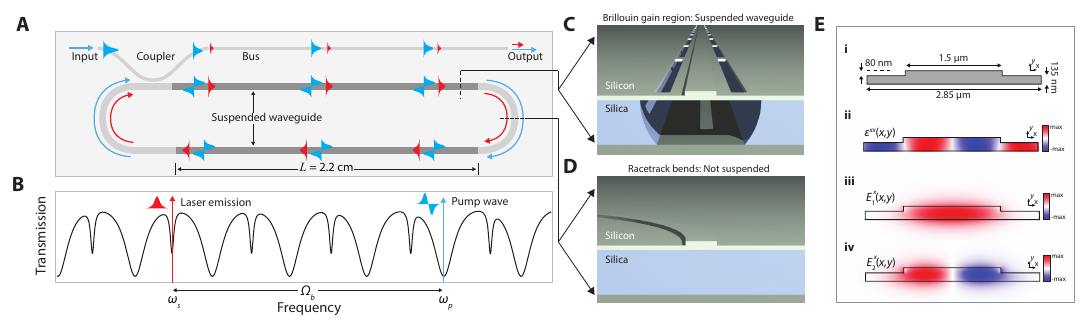}
\caption{Schematic of laser cavity and basic operation. (A) The Brillouin laser consists of a multimode racetrack cavity with two Brillouin-active regions (dark gray). Pump light (blue) is coupled into the antisymmetric spatial mode of the racetrack resonator. Inter-modal Brillouin scattering mediates energy transfer from the pump wave (antisymmetric) to the Stokes wave (symmetric). 
(B) The idealized transmission spectrum for the racetrack cavity.  Narrower (broader) resonant features correspond to the symmetric (antisymmetric) resonances. Brillouin lasing occurs when the resonance condition for the pump (antisymmetric) and Stokes (symmetric) waves are simultaneously satisfied. (C) and (D) illustrate cross sections of the suspended Brillouin-active region and the racetrack bend, respectively. (Ei) Dimensions of Brillouin-active waveguide. (Eii) Strain profile $\epsilon^{xx}(x,y)$ of the 6 GHz Lamb-like acoustic mode that mediates inter-modal scattering. (Eiii) and (Eiv) plot the $x$-directed electric field profiles ($E_x$) of the TE-like symmetric and antisymmetric optical modes, respectively.
}
\label{fig:sims}
\end{figure*}

Within an optical cavity, Brillouin lasing occurs when optical gain from stimulated Brillouin scattering (SBS) overcomes round-trip loss. This nonlinear light-sound coupling is typically strong, overtaking Kerr and Raman interactions in most transparent media. However, the same integrated silicon waveguides that enhance Raman and Kerr nonlinearities tend to produce minuscule Brillouin couplings due to substrate-induced acoustic dissipation \cite{rakichprx}. The recent advent of a class of suspended waveguides---which tightly confine both light and sound---has enabled appreciable nonlinearities through forward stimulated Brillouin scattering \cite{rakichprx,shinnatcomm,roel,Kittlaus2016}. While these suspended structures have produced large optical Brillouin gain \cite{shinnatcomm,roel} and net amplification \cite{van2015net,Kittlaus2016}, innovative strategies are needed to translate Brillouin interactions into silicon laser oscillators \cite{kittlaus2016chip,espinel2017brillouin}.

We demonstrate a Brillouin laser in silicon by leveraging a form of guided-wave forward Brillouin scattering, termed stimulated inter-modal Brillouin scattering (SIMS), which couples light fields guided in distinct optical spatial modes \cite{kittlaus2016chip,kangprl,bahl}. 
Our silicon Brillouin laser system is fabricated from a single-crystal silicon-on-insulator (SOI) wafer (supplementary materials 6.2) \cite{sm}. The laser is composed of a 4.6-cm long racetrack resonator cavity with two extended Brillouin-active gain regions (Fig. 1A). Throughout the device, light is guided by total internal reflection using a ridge waveguide (Fig. 1Ei). This multimode waveguide provides low-loss guidance of both symmetric (red) and antisymmetric (blue) TE-like spatial modes (with respective propagation constants $k_1(\omega)$ and $k_2(\omega)$), yielding two distinct sets of high quality-factor ($Q$) cavity modes with slightly different free spectral ranges (FSR) ($Q_{1} \cong 2.4\times10^6,\rm FSR_1\cong 1.614$ GHz and $ Q_{2} \cong 4\times10^5, \rm FSR_2 \cong 1.570$ GHz, respectively; see supplementary materials 3.7) \cite{sm}. Simulated electric field profiles for these two optical spatial modes are shown in Fig. 1Eiii-iv. To access the cavity modes, we use a directional coupler that couples strongly to the antisymmetric and weakly to the symmetric mode, yielding a characteristic multimode transmission spectrum (Fig. 1B).

Optical gain is supplied by forward inter-modal Brillouin scattering within the Brillouin-active segments (dark gray). These regions are created by removing the oxide undercladding to yield a continuously-suspended waveguide that produces large inter-modal Brillouin gain (Fig. 1C). In addition to low-loss optical modes (Figs. 1Eiii-iv), this structure also supports guidance of a 6 GHz acoustic wave (Fig. 1Eii), which mediates efficient Brillouin coupling between symmetric and antisymmetric optical modes. By contrast, the fixed waveguide bends do not permit acoustic guidance.  The Brillouin-active waveguide structure is identical in design to that described in Ref. \cite{kittlaus2016chip}, which yields a peak inter-modal Brillouin gain coefficient of $G_{\textup{b}} \cong 470 \ {\rm W}^{-1} {\rm m}^{-1}$ at a Brillouin frequency ($\Omega_{\textup{b}}$) of 6.03 GHz with a resonance bandwidth of 13 MHz (full width at half maximum, FWHM). For efficient nonlinear coupling, this scattering process requires that both energy conservation ($\omega_{\textup{p}} = \omega_{\textup{s}}+\Omega_{\textup{b}} $) and phase-matching ($k_{\textup{2}}(\omega_{\textup{p}}) = k_{\textup{1}}(\omega_{\textup{s}}) + q(\Omega_{\textup{b}})$) conditions be satisfied. Here,   $\omega_{\textup{p}}$ and $\omega_{\textup{s}}$ are the respective pump and Stokes frequencies, and $q(\Omega)$ is the wavevector of the acoustic wave.  In inter-modal Brillouin scattering, these conditions produce a form of phase-matched symmetry breaking that decouples the Stokes from the anti-Stokes process, permitting single-sideband amplification \cite{kittlaus2016chip}.

Laser oscillation of the symmetric cavity mode occurs when Brillouin gain matches the round-trip loss, producing coherent laser emission at the Stokes frequency ($\omega_{\rm s}$). These lasing requirements are met by injecting pump light (of power $P_{\rm p}$) into an antisymmetric cavity mode that is separated in frequency from a symmetric cavity mode by the Brillouin frequency (Fig. 1B). Because the FSRs of the two sets of cavity modes differ by 3.1\%, this resonance frequency condition is satisfied by symmetric and antisymmetric cavity mode pairs that occur frequently (every 0.40 nm) across the C-band (from 1530-1565 nm). When this dual-resonance condition is satisfied and the pump power exceeds the threshold power ($P_{\rm p} > P_{\rm th}$), the Stokes field builds from thermal noise (produced by spontaneous Brillouin scattering) to yield appreciable line-narrowing and coherent Stokes emission at frequency $\omega_{\rm s}=\omega_{\rm p}-\Omega_{\textup{b}}$.

Many properties of this system could prove advantageous for scalable and robust integration of Brillouin lasers in complex silicon photonic circuits. Because this laser uses a forward scattering process, it alleviates the need for on-chip isolator and circulator technologies that would otherwise be necessary to integrate traditional Brillouin lasers (which use backward SBS). In addition, as this Brillouin nonlinearity is created through structural control, it is possible to independently engineer a range of characteristics, including Brillouin frequency, acoustic dissipation rate, and Brillouin gain, providing a flexible and robust laser design space. Moreover, the multimode properties of this system eliminate size constraints that are present in backward Brillouin lasers (i.e., FSRs that must correspond to Brillouin frequencies) and provide exceptional control over cascading dynamics (supplementary materials 5.4).  

\begin{figure*}[!tb]
\centering%\vspace{-10pt}
\includegraphics[width=.9\linewidth]{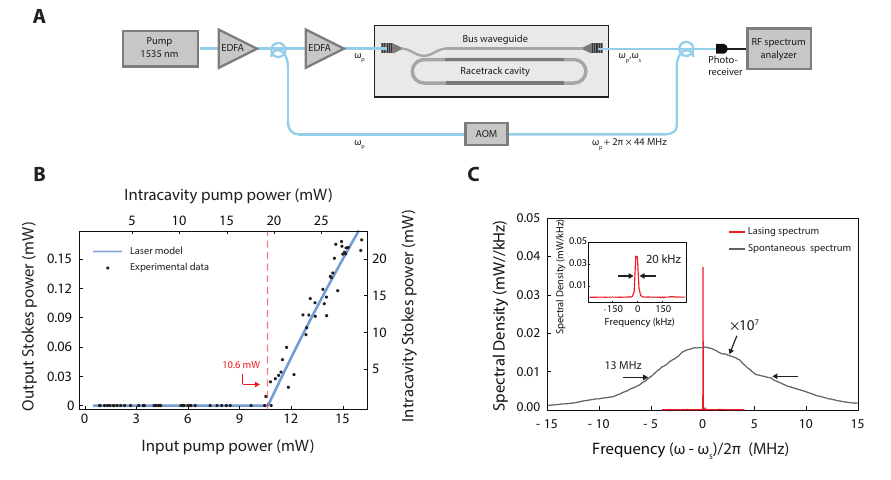}
\caption{Experimental apparatus and laser threshold behavior. (A) Apparatus used for heterodyne spectroscopy. Continuous-wave pump light (Agilent 81600B, linewidth = 13 kHz) used to initiate Brillouin lasing is amplified using an erbium-doped fiber amplifier (EDFA) and coupled on-chip using grating couplers.
Laser light is frequency shifted (+44 MHz) by an acousto-optic modulator (AOM) in a reference arm and combined with the output Stokes light for heterodyne detection.
(B) Theory and experiment for the output laser power vs. input pump power.  Intracavity pump powers are estimated using the transmitted pump power and the detuning from resonance, while intracavity Stokes power is determined from the measured bus Stokes power and comparison with the theoretical model (supplementary materials 4.1) \cite{sm}. (C) Heterodyne spectra of spontaneously scattered Stokes light from a linear waveguide (multiplied by $10^7$) and the linewidth-narrowed intracavity laser spectrum above threshold. 
}
\label{fig:data}
\end{figure*}

Brillouin lasing was investigated by injecting continuous-wave (cw) pump light into an antisymmetric cavity mode while analyzing the emission of Stokes light from a symmetric cavity mode. The power and coherence properties of the emitted laser light were characterized through high-resolution heterodyne spectral analysis (Fig. 2A) \cite{sm}. The threshold and slope efficiency of this laser were quantified by measuring the total emitted Stokes power as a function of pump power (Fig. 2B). These data reveal a threshold on-chip pump power of 10.6 mW, corresponding to an intracavity power of 19 mW. This laser threshold agrees well with the condition for net amplification in Brillouin waveguides of this design \cite{kittlaus2016chip}. Further analysis of these data reveal an on-chip slope efficiency of 3\% (supplementary materials 1.3.2 and 4.1) \cite{sm}.

As the pump power increases, the emitted Stokes light exhibits spectral compression characteristic of laser oscillation. When the emitted Stokes spectrum is broader than the linewidth of optical local oscillator ($\sim 13$ kHz, derived from the same source as the pump wave), the heterodyne microwave spectrum provides an excellent representation of the emitted Stokes linewidth. Fig. 2C compares the Stokes spectrum emitted by the laser (red) with the spontaneous Stokes spectrum emitted from an identical Brillouin-active waveguide segment (gray) in the absence of optical feedback.  We see that optical feedback produces spectral compression by a factor of $\sim10^3$; 
the relatively broad spontaneous Stokes spectrum (FWHM $\cong 13.1$ MHz) is compressed to a resolution-limited value of 20 kHz.

Heterodyne spectral analysis was used to measure the emitted Stokes linewidth below threshold at various Stokes powers (see Fig. 3A and red points of Fig. 3C). A complementary sub-coherence self-heterodyne technique characterizes the laser coherence at higher powers (see Fig. 3B-C and supplementary materials 2) \cite{sm}. Above threshold, the Stokes wave becomes exceptionally coherent with the incident pump field, with an excess phase noise linewidth ($\Delta\nu_{\textup{b}}$) of less than 800 Hz (corresponding to a compression factor of $10^4$).  Due to the three-wave dynamics in this system, this phase noise corresponds directly to the phonon linewidth, revealing phonon linewidth narrowing far below that of the incident pump field.  This behavior represents a marked departure from the linewidth narrowing dynamics conventionally exhibited by Brillouin lasers.

To understand our experimental observations, we derive simple analytical and numerical models that describe the basic spatial and temporal behavior of laser oscillation in this system (see supplementary materials 2) \cite{sm}. Steady-state analysis of the coupled envelope equations reveal that this silicon laser exhibits spatial dynamics (i.e., field evolution along the direction of propagation) that are characteristic of Brillouin lasers.  
Specifically, because the phonon field is spatially heavily damped and the only feedback mechanism is optical, this laser produces optical self-oscillation of the Stokes wave (see supplementary materials 1.1) \cite{sm}.   Building on established treatments of Brillouin laser physics\cite{li2012characterization,debut2000}, a simplified mean-field model was developed to explore the salient features of the temporal dynamics.  This model incorporates parameters that are consistent with the measured resonator and nonlinear waveguide characteristics (see supplementary materials 3.5,3.8) \cite{sm}.

Well above threshold, this model predicts Stokes emission that is highly coherent with the incident pump field, with an excess phase noise linewidth given by 
\begin{align}
\Delta\nu_{\textup{b}} = \frac{\Gamma}{4 \pi \beta^2}(n^{\textup{th}}_{\textup{b}}+n^{\textup{th}}_{\rm s}+1).
\label{eq:lw}
\end{align}
Here, $\Gamma$ is the intrinsic acoustic dissipation rate, $\beta^2$ is the coherently-driven phonon occupation number, $n^{\textup{th}}_{\textup{b}}$ is the thermal occupation number of the phonon field, $n^{\textup{th}}_{\rm s}$ is average thermal occupation number of the symmetric mode of the optical resonator ($n^{\textup{th}}_{\rm b}\approx10^3$ and $n^{\textup{th}}_{\rm s}\approx0$), and the $+1$ is due to vacuum fluctuations. As a result of the three-wave dynamics of this system, the pump-Stokes coherence provides a direct window into the spectrum of the distributed acoustic wave (see supplementary materials 1.3) \cite{sm}, revealing that this regime of Brillouin lasing produces Schawlow-Townes linewidth narrowing of the coherent acoustic field. While closed-form analytical expressions for phase noise are tractable well below and above threshold, stochastic numerical simulations are necessary to model the noise characteristics in the vicinity of laser threshold (see Fig. 3C), revealing good qualitative agreement with our measurements.  

\begin{figure*}[!tb]
\centering%\vspace{-10pt}
\includegraphics[width=.9\linewidth]{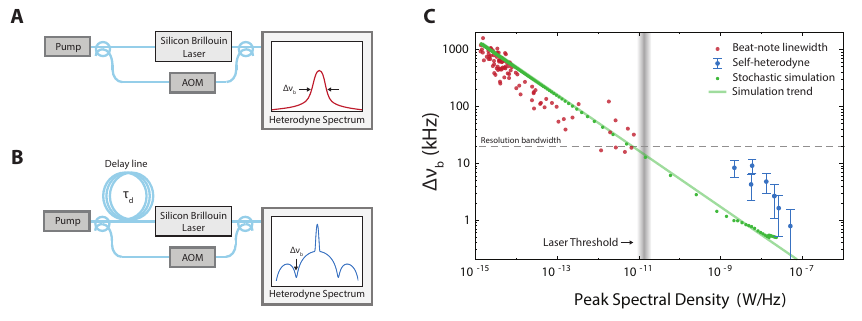}
\caption{Linewidth measurements. (A) Standard heterodyne spectroscopy apparatus to measure the pump-Stokes excess phase noise (or phonon) linewidth ($\Delta \nu_{\rm b}$), given by the FWHM of the heterodyne spectrum. (B) Sub-coherence self-heterodyne apparatus used to probe the phonon dynamics at higher output Stokes powers (supplementary materials 2) \cite{sm}. The phonon linewidth is determined by measuring the fringe contrast or coherence between output Stokes and pump waves (C). (C) Experimental and theoretical comparison of phonon linewidth ($\Delta \nu_{\rm b}$) as a function of peak spectral density. Below threshold we use standard heterodyne spectroscopy (A), (dark red data points). At higher powers, this measurement becomes resolution bandwidth limited (supplementary materials 4.2) \cite{sm}. For this reason, we use the sub-coherence self-heterodyne technique (B), yielding the blue data points (error bars represent the 95\% confidence interval of fits to data, supplementary materials 4.2) \cite{sm}.
}
\label{fig:linewidth}
\end{figure*}

These linewidth narrowing dynamics are distinct from those typically produced in glass-based Brillouin lasers \cite{Smith91,li2012characterization}, which yield Schawlow-Townes optical linewidth narrowing (see supplementary materials 1.6-7) \cite{sm}. In this silicon system, phonon linewidth narrowing arises from an inverted dissipation hierarchy in which the phonon temporal dissipation rate is much smaller than the optical dissipation rates for the pump and Stokes cavity modes (i.e., $\gamma_{\textup{p}},\gamma_{\textup{s}}\gg \Gamma$), in contrast to the temporal dissipation hierarchy conventionally realized in Brillouin lasers ($\gamma_{\textup{p}},\gamma_{\textup{s}}\ll \Gamma$) \cite{debut2000}. As a result, this silicon Brillouin laser simultaneously operates where the spatial acoustic decay length (63 $\rm \mu m$) is far smaller than the optical decay length ($\sim$0.1-1 m), while the intrinsic phonon lifetime (77 ns) exceeds that of the optical fields ($\sim$ 2-12 ns).  This combination of spatial and temporal dynamics is made possible by the unusually large Brillouin coupling in this system ($\sim 10^3 \times$ that of silica fibers) and the disparate velocities of the interacting light and sound waves.

The observed phonon coherence is reminiscent of that produced in optomechanical self-oscillation \cite{grudinin2010phonon,jiang2012high,bahl} (phonon lasing). However, in contrast to phonon lasers, this Brillouin laser does not possess a phonon cavity that permits acoustic feedback necessary for phonon self-oscillation (see supplementary materials 1.7.3-4) \cite{sm}. Here we show that, despite large acoustic spatial damping and lack of phonon feedback (i.e., more than 1000 dB round-trip acoustic propagation loss), optical self-oscillation of the Stokes wave produces linewidth narrowing of the acoustic field, as long as the temporal acoustic dissipation rate is lower than that of the optical fields. In this way, this system is analogous to an extreme limit of singly-resonant optical parametric oscillator physics, with a slow, ballistic, and long-lived idler wave (see supplementary materials 1.7) \cite{sm}. Interestingly, other Brillouin laser systems may have operated near or in this temporal dissipation hierarchy \cite{hill1976cw,abedin2006single,morrison2017compact}. However, these dynamics were not identified. It is due to the stability of this monolithic silicon system that we are able to study this unusual combination of spatial and temporal dynamics.

\subsection{Acknowledgements}
We thank A. D. Stone for discussions regarding the laser dynamics and M. Rooks for assistance with the fabrication. \textbf{Funding:} This work was supported through a seedling grant under the direction of D. Green at DARPA MTO and by the Packard Fellowship for Science and Engineering. N.T.O. acknowledges support from the National Science Foundation Graduate Research Fellowship under grant No. DGE1122492. Z.W. acknowledges support from the National Science Foundation under grant No. 1641069. Any opinion, findings, and conclusions or recommendations expressed in this material are those of the authors and do not necessarily reflect the views of DARPA or the National Science Foundation. \textbf{Author contributions:} P.T.R., N.T.O., and E.A.K. developed device concept and design. N.T.O. and E.A.K. fabricated the devices and performed the experiments. R.O.B. and N.T.O. developed theory with assistance from P.T.R. and Z.W. All authors contributed to the preparation of the manuscript. \textbf{Competing interests:} None declared. \textbf{Data and materials availability:} All data needed to evaluate the conclusions in the paper are present in the paper or the supplementary materials.

\bibliography{cites}

\begin{thebibliography}{30}
\expandafter\ifx\csname natexlab\endcsname\relax\def\natexlab#1{#1}\fi
\expandafter\ifx\csname bibnamefont\endcsname\relax
  \def\bibnamefont#1{#1}\fi
\expandafter\ifx\csname bibfnamefont\endcsname\relax
  \def\bibfnamefont#1{#1}\fi
\expandafter\ifx\csname citenamefont\endcsname\relax
  \def\citenamefont#1{#1}\fi
\expandafter\ifx\csname url\endcsname\relax
  \def\url#1{\texttt{#1}}\fi
\expandafter\ifx\csname urlprefix\endcsname\relax\def\urlprefix{URL }\fi
\providecommand{\bibinfo}[2]{#2}
\providecommand{\eprint}[2][]{\url{#2}}

\bibitem[{\citenamefont{Reed et~al.}(2010)\citenamefont{Reed, Mashanovich,
  Gardes, and Thomson}}]{Reed2010}
\bibinfo{author}{\bibfnamefont{G.~T.} \bibnamefont{Reed}},
  \bibinfo{author}{\bibfnamefont{G.}~\bibnamefont{Mashanovich}},
  \bibinfo{author}{\bibfnamefont{F.~Y.} \bibnamefont{Gardes}},
  \bibnamefont{and} \bibinfo{author}{\bibfnamefont{D.~J.}
  \bibnamefont{Thomson}}, \bibinfo{journal}{Nat. Photonics}
  \textbf{\bibinfo{volume}{4}}, \bibinfo{pages}{518} (\bibinfo{year}{2010}),
  ISSN \bibinfo{issn}{1749-4885},
  \urlprefix\url{http://dx.doi.org/10.1038/nphoton.2010.179}.

\bibitem[{\citenamefont{Liang and Bowers}(2010)}]{liang2010recent}
\bibinfo{author}{\bibfnamefont{D.}~\bibnamefont{Liang}} \bibnamefont{and}
  \bibinfo{author}{\bibfnamefont{J.~E.} \bibnamefont{Bowers}},
  \bibinfo{journal}{Nat. Photonics} \textbf{\bibinfo{volume}{4}},
  \bibinfo{pages}{511} (\bibinfo{year}{2010}).

\bibitem[{\citenamefont{Leuthold et~al.}(2010)\citenamefont{Leuthold, Koos, and
  Freude}}]{leuthold2010nonlinear}
\bibinfo{author}{\bibfnamefont{J.}~\bibnamefont{Leuthold}},
  \bibinfo{author}{\bibfnamefont{C.}~\bibnamefont{Koos}}, \bibnamefont{and}
  \bibinfo{author}{\bibfnamefont{W.}~\bibnamefont{Freude}},
  \bibinfo{journal}{Nat. Photonics} \textbf{\bibinfo{volume}{4}},
  \bibinfo{pages}{535} (\bibinfo{year}{2010}).

\bibitem[{\citenamefont{Alduino and Paniccia}(2007)}]{alduino2007interconnects}
\bibinfo{author}{\bibfnamefont{A.}~\bibnamefont{Alduino}} \bibnamefont{and}
  \bibinfo{author}{\bibfnamefont{M.}~\bibnamefont{Paniccia}},
  \bibinfo{journal}{Nat. Photonics} \textbf{\bibinfo{volume}{1}},
  \bibinfo{pages}{153} (\bibinfo{year}{2007}).

\bibitem[{\citenamefont{Vos et~al.}(2007)\citenamefont{Vos, Bartolozzi,
  Schacht, Bienstman, and Baets}}]{devos07}
\bibinfo{author}{\bibfnamefont{K.~D.} \bibnamefont{Vos}},
  \bibinfo{author}{\bibfnamefont{I.}~\bibnamefont{Bartolozzi}},
  \bibinfo{author}{\bibfnamefont{E.}~\bibnamefont{Schacht}},
  \bibinfo{author}{\bibfnamefont{P.}~\bibnamefont{Bienstman}},
  \bibnamefont{and} \bibinfo{author}{\bibfnamefont{R.}~\bibnamefont{Baets}},
  \bibinfo{journal}{Opt. Express} \textbf{\bibinfo{volume}{15}},
  \bibinfo{pages}{7610} (\bibinfo{year}{2007}),
  \urlprefix\url{http://www.opticsexpress.org/abstract.cfm?URI=oe-15-12-7610}.

\bibitem[{\citenamefont{Boyraz and Jalali}(2004)}]{boyraz2004demonstration}
\bibinfo{author}{\bibfnamefont{O.}~\bibnamefont{Boyraz}} \bibnamefont{and}
  \bibinfo{author}{\bibfnamefont{B.}~\bibnamefont{Jalali}},
  \bibinfo{journal}{Opt. Express} \textbf{\bibinfo{volume}{12}},
  \bibinfo{pages}{5269} (\bibinfo{year}{2004}).

\bibitem[{\citenamefont{Rong et~al.}(2005)\citenamefont{Rong, Jones, Liu,
  Cohen, Hak, Fang, and Paniccia}}]{rong2005continuous}
\bibinfo{author}{\bibfnamefont{H.}~\bibnamefont{Rong}},
  \bibinfo{author}{\bibfnamefont{R.}~\bibnamefont{Jones}},
  \bibinfo{author}{\bibfnamefont{A.}~\bibnamefont{Liu}},
  \bibinfo{author}{\bibfnamefont{O.}~\bibnamefont{Cohen}},
  \bibinfo{author}{\bibfnamefont{D.}~\bibnamefont{Hak}},
  \bibinfo{author}{\bibfnamefont{A.}~\bibnamefont{Fang}}, \bibnamefont{and}
  \bibinfo{author}{\bibfnamefont{M.}~\bibnamefont{Paniccia}},
  \bibinfo{journal}{Nature} \textbf{\bibinfo{volume}{433}},
  \bibinfo{pages}{725} (\bibinfo{year}{2005}).

\bibitem[{\citenamefont{Takuma and Jennings}(1964)}]{takuma1964stimulated}
\bibinfo{author}{\bibfnamefont{H.}~\bibnamefont{Takuma}} \bibnamefont{and}
  \bibinfo{author}{\bibfnamefont{D.}~\bibnamefont{Jennings}},
  \bibinfo{journal}{Appl. Phys. Lett.} \textbf{\bibinfo{volume}{5}},
  \bibinfo{pages}{239} (\bibinfo{year}{1964}).

\bibitem[{\citenamefont{Lecoeuche et~al.}(1998)\citenamefont{Lecoeuche, Webb,
  Pannell, and Jackson}}]{lecoeuche1998brillouin}
\bibinfo{author}{\bibfnamefont{V.}~\bibnamefont{Lecoeuche}},
  \bibinfo{author}{\bibfnamefont{D.~J.} \bibnamefont{Webb}},
  \bibinfo{author}{\bibfnamefont{C.~N.} \bibnamefont{Pannell}},
  \bibnamefont{and} \bibinfo{author}{\bibfnamefont{D.~A.}
  \bibnamefont{Jackson}}, \bibinfo{journal}{Opt. Commun.}
  \textbf{\bibinfo{volume}{152}}, \bibinfo{pages}{263} (\bibinfo{year}{1998}).

\bibitem[{\citenamefont{Smith et~al.}(1991)\citenamefont{Smith, Zarinetchi, and
  Ezekiel}}]{Smith91}
\bibinfo{author}{\bibfnamefont{S.~P.} \bibnamefont{Smith}},
  \bibinfo{author}{\bibfnamefont{F.}~\bibnamefont{Zarinetchi}},
  \bibnamefont{and} \bibinfo{author}{\bibfnamefont{S.}~\bibnamefont{Ezekiel}},
  \bibinfo{journal}{Opt. Lett.} \textbf{\bibinfo{volume}{16}},
  \bibinfo{pages}{393} (\bibinfo{year}{1991}),
  \urlprefix\url{http://ol.osa.org/abstract.cfm?URI=ol-16-6-393}.

\bibitem[{\citenamefont{Loh et~al.}(2015)\citenamefont{Loh, Green, Baynes,
  Cole, Quinlan, Lee, Vahala, Papp, and Diddams}}]{Loh15}
\bibinfo{author}{\bibfnamefont{W.}~\bibnamefont{Loh}},
  \bibinfo{author}{\bibfnamefont{A.~A.~S.} \bibnamefont{Green}},
  \bibinfo{author}{\bibfnamefont{F.~N.} \bibnamefont{Baynes}},
  \bibinfo{author}{\bibfnamefont{D.~C.} \bibnamefont{Cole}},
  \bibinfo{author}{\bibfnamefont{F.~J.} \bibnamefont{Quinlan}},
  \bibinfo{author}{\bibfnamefont{H.}~\bibnamefont{Lee}},
  \bibinfo{author}{\bibfnamefont{K.~J.} \bibnamefont{Vahala}},
  \bibinfo{author}{\bibfnamefont{S.~B.} \bibnamefont{Papp}}, \bibnamefont{and}
  \bibinfo{author}{\bibfnamefont{S.~A.} \bibnamefont{Diddams}},
  \bibinfo{journal}{Optica} \textbf{\bibinfo{volume}{2}}, \bibinfo{pages}{225}
  (\bibinfo{year}{2015}),
  \urlprefix\url{http://www.osapublishing.org/optica/abstract.cfm?URI=optica-2-3-225}.

\bibitem[{\citenamefont{Kabakova et~al.}(2013)\citenamefont{Kabakova, Pant,
  Choi, Debbarma, Luther-Davies, Madden, and Eggleton}}]{kabakova2013narrow}
\bibinfo{author}{\bibfnamefont{I.~V.} \bibnamefont{Kabakova}},
  \bibinfo{author}{\bibfnamefont{R.}~\bibnamefont{Pant}},
  \bibinfo{author}{\bibfnamefont{D.-Y.} \bibnamefont{Choi}},
  \bibinfo{author}{\bibfnamefont{S.}~\bibnamefont{Debbarma}},
  \bibinfo{author}{\bibfnamefont{B.}~\bibnamefont{Luther-Davies}},
  \bibinfo{author}{\bibfnamefont{S.~J.} \bibnamefont{Madden}},
  \bibnamefont{and} \bibinfo{author}{\bibfnamefont{B.~J.}
  \bibnamefont{Eggleton}}, \bibinfo{journal}{Opt. Lett.}
  \textbf{\bibinfo{volume}{38}}, \bibinfo{pages}{3208} (\bibinfo{year}{2013}).

\bibitem[{\citenamefont{Li et~al.}(2017)\citenamefont{Li, Suh, and
  Vahala}}]{li2017microresonator}
\bibinfo{author}{\bibfnamefont{J.}~\bibnamefont{Li}},
  \bibinfo{author}{\bibfnamefont{M.-G.} \bibnamefont{Suh}}, \bibnamefont{and}
  \bibinfo{author}{\bibfnamefont{K.}~\bibnamefont{Vahala}},
  \bibinfo{journal}{Optica} \textbf{\bibinfo{volume}{4}}, \bibinfo{pages}{346}
  (\bibinfo{year}{2017}).

\bibitem[{\citenamefont{Rakich et~al.}(2012)\citenamefont{Rakich, Reinke,
  Camacho, Davids, and Wang}}]{rakichprx}
\bibinfo{author}{\bibfnamefont{P.~T.} \bibnamefont{Rakich}},
  \bibinfo{author}{\bibfnamefont{C.}~\bibnamefont{Reinke}},
  \bibinfo{author}{\bibfnamefont{R.}~\bibnamefont{Camacho}},
  \bibinfo{author}{\bibfnamefont{P.}~\bibnamefont{Davids}}, \bibnamefont{and}
  \bibinfo{author}{\bibfnamefont{Z.}~\bibnamefont{Wang}},
  \bibinfo{journal}{Phys. Rev. X} \textbf{\bibinfo{volume}{2}},
  \bibinfo{pages}{011008} (\bibinfo{year}{2012}),
  \urlprefix\url{http://link.aps.org/doi/10.1103/PhysRevX.2.011008}.

\bibitem[{\citenamefont{Shin et~al.}(2013)\citenamefont{Shin, Qiu, Jarecki,
  Cox, Olsson, Starbuck, Wang, and Rakich}}]{shinnatcomm}
\bibinfo{author}{\bibfnamefont{H.}~\bibnamefont{Shin}},
  \bibinfo{author}{\bibfnamefont{W.}~\bibnamefont{Qiu}},
  \bibinfo{author}{\bibfnamefont{R.}~\bibnamefont{Jarecki}},
  \bibinfo{author}{\bibfnamefont{J.~A.} \bibnamefont{Cox}},
  \bibinfo{author}{\bibfnamefont{R.~H.} \bibnamefont{Olsson}},
  \bibinfo{author}{\bibfnamefont{A.}~\bibnamefont{Starbuck}},
  \bibinfo{author}{\bibfnamefont{Z.}~\bibnamefont{Wang}}, \bibnamefont{and}
  \bibinfo{author}{\bibfnamefont{P.~T.} \bibnamefont{Rakich}},
  \bibinfo{journal}{Nat. Commun.} \textbf{\bibinfo{volume}{4}},
  \bibinfo{pages}{1944} (\bibinfo{year}{2013}),
  \urlprefix\url{http://dx.doi.org/10.1038/ncomms2943}.

\bibitem[{\citenamefont{Van~Laer et~al.}(2015)\citenamefont{Van~Laer, Kuyken,
  Van~Thourhout, and Baets}}]{roel}
\bibinfo{author}{\bibfnamefont{R.}~\bibnamefont{Van~Laer}},
  \bibinfo{author}{\bibfnamefont{B.}~\bibnamefont{Kuyken}},
  \bibinfo{author}{\bibfnamefont{D.}~\bibnamefont{Van~Thourhout}},
  \bibnamefont{and} \bibinfo{author}{\bibfnamefont{R.}~\bibnamefont{Baets}},
  \bibinfo{journal}{Nat. Photonics} \textbf{\bibinfo{volume}{9}},
  \bibinfo{pages}{199} (\bibinfo{year}{2015}), ISSN \bibinfo{issn}{1749-4885},
  \urlprefix\url{http://dx.doi.org/10.1038/nphoton.2015.11}.

\bibitem[{\citenamefont{Kittlaus et~al.}(2016)\citenamefont{Kittlaus, Shin, and
  Rakich}}]{Kittlaus2016}
\bibinfo{author}{\bibfnamefont{E.~A.} \bibnamefont{Kittlaus}},
  \bibinfo{author}{\bibfnamefont{H.}~\bibnamefont{Shin}}, \bibnamefont{and}
  \bibinfo{author}{\bibfnamefont{P.~T.} \bibnamefont{Rakich}},
  \bibinfo{journal}{Nat. Photonics} \textbf{\bibinfo{volume}{10}},
  \bibinfo{pages}{463} (\bibinfo{year}{2016}), ISSN \bibinfo{issn}{1749-4885},
  \urlprefix\url{http://dx.doi.org/10.1038/nphoton.2016.112}.

\bibitem[{\citenamefont{Laer et~al.}(2015)\citenamefont{Laer, Bazin, Kuyken,
  Baets, and Thourhout}}]{van2015net}
\bibinfo{author}{\bibfnamefont{R.~V.} \bibnamefont{Laer}},
  \bibinfo{author}{\bibfnamefont{A.}~\bibnamefont{Bazin}},
  \bibinfo{author}{\bibfnamefont{B.}~\bibnamefont{Kuyken}},
  \bibinfo{author}{\bibfnamefont{R.}~\bibnamefont{Baets}}, \bibnamefont{and}
  \bibinfo{author}{\bibfnamefont{D.~V.} \bibnamefont{Thourhout}},
  \bibinfo{journal}{New J. Phys.} \textbf{\bibinfo{volume}{17}},
  \bibinfo{pages}{115005} (\bibinfo{year}{2015}),
  \urlprefix\url{http://stacks.iop.org/1367-2630/17/i=11/a=115005}.

\bibitem[{\citenamefont{Kittlaus et~al.}(2017)\citenamefont{Kittlaus,
  Otterstrom, and Rakich}}]{kittlaus2016chip}
\bibinfo{author}{\bibfnamefont{E.~A.} \bibnamefont{Kittlaus}},
  \bibinfo{author}{\bibfnamefont{N.~T.} \bibnamefont{Otterstrom}},
  \bibnamefont{and} \bibinfo{author}{\bibfnamefont{P.~T.}
  \bibnamefont{Rakich}}, \bibinfo{journal}{Nat. Commun.}
  \textbf{\bibinfo{volume}{8}}, \bibinfo{pages}{15819} (\bibinfo{year}{2017}).

\bibitem[{\citenamefont{Espinel et~al.}(2017)\citenamefont{Espinel, Santos,
  Luiz, Alegre, and Wiederhecker}}]{espinel2017brillouin}
\bibinfo{author}{\bibfnamefont{Y.~A.} \bibnamefont{Espinel}},
  \bibinfo{author}{\bibfnamefont{F.~G.} \bibnamefont{Santos}},
  \bibinfo{author}{\bibfnamefont{G.~O.} \bibnamefont{Luiz}},
  \bibinfo{author}{\bibfnamefont{T.~M.} \bibnamefont{Alegre}},
  \bibnamefont{and} \bibinfo{author}{\bibfnamefont{G.~S.}
  \bibnamefont{Wiederhecker}}, \bibinfo{journal}{Sci. Rep.}
  \textbf{\bibinfo{volume}{7}}, \bibinfo{pages}{43423} (\bibinfo{year}{2017}).

\bibitem[{\citenamefont{Kang et~al.}(2010)\citenamefont{Kang, Brenn, and
  St.J.~Russell}}]{kangprl}
\bibinfo{author}{\bibfnamefont{M.~S.} \bibnamefont{Kang}},
  \bibinfo{author}{\bibfnamefont{A.}~\bibnamefont{Brenn}}, \bibnamefont{and}
  \bibinfo{author}{\bibfnamefont{P.}~\bibnamefont{St.J.~Russell}},
  \bibinfo{journal}{Phys. Rev. Lett.} \textbf{\bibinfo{volume}{105}},
  \bibinfo{pages}{153901} (\bibinfo{year}{2010}),
  \urlprefix\url{http://link.aps.org/doi/10.1103/PhysRevLett.105.153901}.

\bibitem[{\citenamefont{Bahl et~al.}(2011)\citenamefont{Bahl, Zehnpfennig,
  Tomes, and Carmon}}]{bahl}
\bibinfo{author}{\bibfnamefont{G.}~\bibnamefont{Bahl}},
  \bibinfo{author}{\bibfnamefont{J.}~\bibnamefont{Zehnpfennig}},
  \bibinfo{author}{\bibfnamefont{M.}~\bibnamefont{Tomes}}, \bibnamefont{and}
  \bibinfo{author}{\bibfnamefont{T.}~\bibnamefont{Carmon}},
  \bibinfo{journal}{Nat. Commun.} \textbf{\bibinfo{volume}{2}},
  \bibinfo{pages}{403} (\bibinfo{year}{2011}).

\bibitem[{\citenamefont{Materials and methods are available as~supplementary
  materials}(2018)}]{sm}
\bibinfo{author}{\bibnamefont{Materials}} \bibnamefont{and}
  \bibinfo{author}{\bibnamefont{methods are available as~supplementary
  materials}} (\bibinfo{year}{2018}).

\bibitem[{\citenamefont{Li et~al.}(2012)\citenamefont{Li, Lee, Chen, and
  Vahala}}]{li2012characterization}
\bibinfo{author}{\bibfnamefont{J.}~\bibnamefont{Li}},
  \bibinfo{author}{\bibfnamefont{H.}~\bibnamefont{Lee}},
  \bibinfo{author}{\bibfnamefont{T.}~\bibnamefont{Chen}}, \bibnamefont{and}
  \bibinfo{author}{\bibfnamefont{K.~J.} \bibnamefont{Vahala}},
  \bibinfo{journal}{Opt. Express} \textbf{\bibinfo{volume}{20}},
  \bibinfo{pages}{20170} (\bibinfo{year}{2012}).

\bibitem[{\citenamefont{Debut et~al.}(2000)\citenamefont{Debut, Randoux, and
  Zemmouri}}]{debut2000}
\bibinfo{author}{\bibfnamefont{A.}~\bibnamefont{Debut}},
  \bibinfo{author}{\bibfnamefont{S.}~\bibnamefont{Randoux}}, \bibnamefont{and}
  \bibinfo{author}{\bibfnamefont{J.}~\bibnamefont{Zemmouri}},
  \bibinfo{journal}{Phys. Rev. A} \textbf{\bibinfo{volume}{62}},
  \bibinfo{pages}{023803} (\bibinfo{year}{2000}),
  \urlprefix\url{https://link.aps.org/doi/10.1103/PhysRevA.62.023803}.

\bibitem[{\citenamefont{Grudinin et~al.}(2010)\citenamefont{Grudinin, Lee,
  Painter, and Vahala}}]{grudinin2010phonon}
\bibinfo{author}{\bibfnamefont{I.~S.} \bibnamefont{Grudinin}},
  \bibinfo{author}{\bibfnamefont{H.}~\bibnamefont{Lee}},
  \bibinfo{author}{\bibfnamefont{O.}~\bibnamefont{Painter}}, \bibnamefont{and}
  \bibinfo{author}{\bibfnamefont{K.~J.} \bibnamefont{Vahala}},
  \bibinfo{journal}{Phys. Rev. Lett.} \textbf{\bibinfo{volume}{104}},
  \bibinfo{pages}{083901} (\bibinfo{year}{2010}).

\bibitem[{\citenamefont{Jiang et~al.}(2012)\citenamefont{Jiang, Lu, Zhang, and
  Lin}}]{jiang2012high}
\bibinfo{author}{\bibfnamefont{W.~C.} \bibnamefont{Jiang}},
  \bibinfo{author}{\bibfnamefont{X.}~\bibnamefont{Lu}},
  \bibinfo{author}{\bibfnamefont{J.}~\bibnamefont{Zhang}}, \bibnamefont{and}
  \bibinfo{author}{\bibfnamefont{Q.}~\bibnamefont{Lin}}, \bibinfo{journal}{Opt.
  Express} \textbf{\bibinfo{volume}{20}}, \bibinfo{pages}{15991}
  (\bibinfo{year}{2012}).

\bibitem[{\citenamefont{Hill et~al.}(1976)\citenamefont{Hill, Kawasaki, and
  Johnson}}]{hill1976cw}
\bibinfo{author}{\bibfnamefont{K.}~\bibnamefont{Hill}},
  \bibinfo{author}{\bibfnamefont{B.}~\bibnamefont{Kawasaki}}, \bibnamefont{and}
  \bibinfo{author}{\bibfnamefont{D.}~\bibnamefont{Johnson}},
  \bibinfo{journal}{Appl. Phys. Lett.} \textbf{\bibinfo{volume}{28}},
  \bibinfo{pages}{608} (\bibinfo{year}{1976}).

\bibitem[{\citenamefont{Abedin}(2006)}]{abedin2006single}
\bibinfo{author}{\bibfnamefont{K.~S.} \bibnamefont{Abedin}},
  \bibinfo{journal}{Opt. Express} \textbf{\bibinfo{volume}{14}},
  \bibinfo{pages}{4037} (\bibinfo{year}{2006}).

\bibitem[{\citenamefont{Morrison et~al.}(2017)\citenamefont{Morrison,
  Casas-Bedoya, Ren, Vu, Liu, Zarifi, Nguyen, Choi, Marpaung, Madden
  et~al.}}]{morrison2017compact}
\bibinfo{author}{\bibfnamefont{B.}~\bibnamefont{Morrison}},
  \bibinfo{author}{\bibfnamefont{A.}~\bibnamefont{Casas-Bedoya}},
  \bibinfo{author}{\bibfnamefont{G.}~\bibnamefont{Ren}},
  \bibinfo{author}{\bibfnamefont{K.}~\bibnamefont{Vu}},
  \bibinfo{author}{\bibfnamefont{Y.}~\bibnamefont{Liu}},
  \bibinfo{author}{\bibfnamefont{A.}~\bibnamefont{Zarifi}},
  \bibinfo{author}{\bibfnamefont{T.~G.} \bibnamefont{Nguyen}},
  \bibinfo{author}{\bibfnamefont{D.-Y.} \bibnamefont{Choi}},
  \bibinfo{author}{\bibfnamefont{D.}~\bibnamefont{Marpaung}},
  \bibinfo{author}{\bibfnamefont{S.~J.} \bibnamefont{Madden}},
  \bibnamefont{et~al.}, \bibinfo{journal}{Optica} \textbf{\bibinfo{volume}{4}},
  \bibinfo{pages}{847} (\bibinfo{year}{2017}).

\end{thebibliography}
\newpage\null\thispagestyle{empty}\newpage


\section*{Supplementary material}

# Supplementary Materials for
## A silicon Brillouin laser

Nils T. Otterstrom,[†] Ryan O. Behunin, Eric A. Kittlaus,
Zheng Wang, Peter T. Rakich[†]

[†] Corresponding author emails: nils.otterstrom@yale.edu, peter.rakich@yale.edu

**This PDF file includes:**

# Contents



# 1   Model of inter-modal Brillouin laser dynamics

In this section, we present a simple model that describes the salient features of the dynamics of our inter-modal Brillouin laser. In what follows, we first explore the spatial dynamics of this Brillouin laser. This approach allows us to identify the origin of laser oscillation in this system and derive the threshold condition. We also show that the spatial behavior of this system is identical to conventional Brillouin lasers. After analyzing the spatial dynamics, we make a series of approximations to develop a mean-field model of the laser physics — a reduction that is adequate to capture the essential aspects of the temporal dynamics. After discussing the validity of this model, we corroborate the laser threshold and then derive the slope efficiency and noise properties. In particular, we show that in contrast to the temporal dynamics conventionally exhibited in Brillouin lasers, optical self-oscillation in this system causes the phononic degrees of freedom to undergo strong spectral compression.

## 1.1   Spatial dynamics and origin of laser oscillation

Stimulated Brillouin scattering is a three-wave interaction that produces spatial stimulated optical gain. These characteristic Brillouin dynamics arise because the spatial decay rate of the phonon field is much larger than the spatial optical decay rates. In this section, we show that this silicon Brillouin laser exhibits spatial dynamics which are consistent with other Brillouin lasers. Due to the large acoustic spatial decay rate, the phonon field can be eliminated from the spatial equations of motion, resulting in exponential spatial amplification for the Stokes field while the phonon field strength is given by the local optical beat note between pump and Stokes fields (*31, 32*). We also derive the threshold condition and show that laser oscillation occurs when round-trip optical Brillouin gain balances round-trip optical loss of the Stokes wave.

We begin our analysis from the envelope equations of motion (*33, 34*) given by

$$\dot{A}_\text{p} + i\omega_\text{p} A_\text{p} + v_\text{g,p} A'_\text{p} + \frac{1}{2} v_\text{g,p} \alpha_\text{p} A_\text{p} = -i\tilde{g} A_\text{s} B \tag{1}$$

$$\dot{A}_\text{s} + i\omega_\text{s} A_\text{s} + v_\text{g,s} A'_\text{s} + \frac{1}{2} v_\text{g,s} \alpha_\text{s} A_\text{s} = -i\tilde{g}^* A_\text{p} B^\dagger \tag{2}$$

$$\dot{B} + (i\Omega_\text{b} + \frac{1}{2}\Gamma)B - v_\text{g,b} B' = -i\tilde{g}^* A_\text{p} A_\text{s}^\dagger. \tag{3}$$

Here, $\tilde{g}$ is the distributed Brillouin coupling, $v_\text{g,p}$ ($v_\text{g,s}$) is the group velocity of the pump (Stokes) field, $v_\text{g,b}$ is the group velocity of the phonon field, and $\alpha_\text{p}$ ($\alpha_\text{s}$) is the spatial decay rate of the pump (Stokes) field in the inter-modal waveguide. To simplify our analysis of the threshold condition, we move to the rotating frame, assume an undepleted pump, and solve for the steady-state spatial dynamics. The resulting coupled slowly-varying envelope equations are

$$v_\text{g,s} A'_\text{s} = -i\tilde{g}^* A_\text{p} B^\dagger - \frac{1}{2} v_\text{g,s} \alpha_\text{s} A_\text{s} \tag{4}$$

$$-v_\text{g,b} B' = -i\tilde{g}^* A_\text{p} A_\text{s}^\dagger - \frac{1}{2}\Gamma B. \tag{5}$$

From these governing equations, the formal solution of $B(z)$ is

$$B(z) = \frac{-i\tilde{g}^* A_\text{p}}{v_\text{g,b}} \int_z^L dz' A_\text{s}^\dagger(z') e^{\frac{\Gamma(z-z')}{2v_\text{g,b}}}. \tag{6}$$

This expression greatly simplifies by recognizing that the Stokes wave varies very slowly compared to the spatial decay length of the phonon field. This fact allows us to pull the Stokes envelope out of the integral, resulting in a simple spatial solution of the phonon field:

$$B(z) = \frac{-2i\tilde{g}^*}{\Gamma} A_{\mathrm{p}} A_{\mathrm{s}}^{\dagger}(z)(1 - e^{\frac{-\Gamma(L-z)}{2v_{\mathrm{g,b}}}}). \tag{7}$$

For $(L - z) \gg \frac{v_{\mathrm{g,b}}}{\Gamma}$ (which is very well-satisfied in our device), this expression reduces to

$$B(z) = \frac{-2i\tilde{g}^*}{\Gamma} A_{\mathrm{p}} A_{\mathrm{s}}^{\dagger}(z). \tag{8}$$

From Eq. 8, we see that the phonon field strength is given by the local pump-Stokes beat note. This aspect of the spatial dynamics is common to practically all Brillouin systems and thoroughly discussed in standard nonlinear optics textbooks (see (*31, 32*)). We can now insert our solution for $B(z)$ into the equation of motion for the Stokes field, which becomes

$$A_{\mathrm{s}}' = \frac{2|\tilde{g}|^2 |A_{\mathrm{p}}|^2}{v_{\mathrm{g,s}}\Gamma} A_{\mathrm{s}} - \frac{\alpha_{\mathrm{s}}}{2} A_{\mathrm{s}}. \tag{9}$$

Substituting the known expression for $G_{\mathrm{b}}$ (*34*) and solving Eq. 9, we find the following solution for the Stokes field:

$$A_{\mathrm{s}} = C_1 e^{(\frac{G_{\mathrm{b}}P_{\mathrm{P}}}{2} - \frac{\alpha_{\mathrm{s}}}{2})z}. \tag{10}$$

We next apply the boundary conditions to determine the value of $C_1$. The Stokes cavity is formed by fashioning a multi-mode waveguide into a racetrack ring configuration with an input/output coupler. Consistent with the physical parameters of the device under study, here we apply no acoustic feedback (Section 1.7.4). The boundary condition is therefore expressed as

$$A_{\mathrm{s}}(0) = -i\mu A_{\mathrm{s,in}} + r A_{\mathrm{s}}(L). \tag{11}$$

Here, $|\mu|^2$ is the coupling into/out of the ring and $r$ parameterizes the fraction of light that feeds back into the ring after the coupler ($r = \sqrt{1 - \mu^2}$). In our laser, there is no input Stokes light, and thus in this simple analysis, $A_{\mathrm{s,in}}$ serves as a proxy for the noise that initiates laser oscillation. Applying this boundary condition, we find that the solution for the Stokes field envelope is given by

$$A_{\mathrm{s}}(z) = \frac{-i\mu A_{\mathrm{s,in}}}{1 - r e^{\frac{(G_{\mathrm{b}}P_{\mathrm{P}} - \alpha_{\mathrm{s}})L}{2}}} e^{\frac{(G_{\mathrm{b}}P_{\mathrm{P}} - \alpha_{\mathrm{s}})z}{2}}. \tag{12}$$

This is the main result of this section. Here we wish to make a couple of important observations. From Eq. 12, we see that the field diverges when $G_{\mathrm{b}}P_{\mathrm{P}} = \alpha_{\mathrm{s}} + \frac{2}{L}\log\frac{1}{r}$. In other words, the Stokes field experiences a pole on the real axis (a lossless state) when round-trip Brillouin gain matches the optical round-trip loss; a condition which pushes the system into optical self-oscillation. This is the threshold condition for Brillouin lasing (*32*).

From Eq. 8, we note that above threshold, optical self-oscillation also dramatically increases the phonon field strength. Since the phonon field is spatially heavily damped, the spatial dynamics are determined by the local optical driving force, which radically increases in magnitude above threshold. Thus, because the phonon field mediates coupling between the pump and Stokes waves, phonon emission can be viewed as a byproduct of Stokes laser-oscillation (much like the idler in a singly-resonant OPO, see Section 1.7). Thus, in general, Brillouin lasers require only an optical cavity for the Stokes field; no phonon feedback is necessary for laser oscillation to occur.

In subsequent sections, we will see that even though the phonon field is not responsible for the feedback and oscillation, it plays a crucial role in the linewidth narrowing dynamics of this silicon Brillouin laser because it is the carrier with the longest temporal memory.

## 1.2   Mean-field analysis

We next develop a simple mean-field (*35*) model that captures the basic temporal and noise dynamics of this laser. This model can be derived by again starting from the slowly-varying envelope dynamics of these fields (*33, 34*) given by

$$\dot{A}_{\rm p} + i\omega_{\rm p}A_{\rm p} + v_{\rm g,p}A'_{\rm p} + \frac{1}{2}v_{\rm g,p}\alpha_{\rm p}A_{\rm p} = -i\tilde{g}A_{\rm s}B + \frac{1}{\sqrt{L}}\eta_{\rm p}(z,t) \tag{13}$$

$$\dot{A}_{\rm s} + i\omega_{\rm s}A_{\rm s} + v_{\rm g,s}A'_{\rm s} + \frac{1}{2}v_{\rm g,s}\alpha_{\rm s}A_{\rm s} = -i\tilde{g}^*A_{\rm p}B^\dagger + \frac{1}{\sqrt{L}}\eta_{\rm s}(z,t) \tag{14}$$

$$\dot{B} + (i\Omega_{\rm b} + \frac{1}{2}\Gamma)B + v_{\rm g,b}B' = -i\tilde{g}^*A_{\rm p}A_{\rm s}^\dagger + \frac{1}{\sqrt{L}}\xi(z,t). \tag{15}$$

Here, we have added an explicit spatial dependence for the Langevin forces, which are assumed to be locally correlated in space, e.g. $\langle \xi^\dagger(t,z)\xi(t',z')\rangle = L\Gamma n_{\rm b}^{\rm th}\delta(t-t')\delta(z-z')$ and similarly for the other noise correlation functions.

The dynamics of the mean optical and acoustic fields can be obtained by averaging each envelope over the length of the ring $L$. For example, the mean pump field is given by

$$\bar{A}_{\rm p} \equiv \frac{1}{L}\int_0^L dz\, A_{\rm p}. \tag{16}$$

Applying this averaging procedure to the envelope equation for the pump we find

$$\dot{\bar{A}}_{\rm p} + i\omega_{\rm p}\bar{A}_{\rm p} + \frac{v_{\rm g,p}}{L}(A_{\rm p}(L) - A_{\rm p}(0)) + \frac{1}{2}v_{\rm g,p}\alpha_{\rm p}\bar{A}_{\rm p} = -i\tilde{g}\overline{A_{\rm s}B} + \frac{1}{\sqrt{L}}\bar{\eta}_{\rm p}. \tag{17}$$

This equation can be further simplified by assuming that the envelopes change very little over a round trip, and that the pump light in the ring obeys the boundary condition $A_{\rm p}(0) = r_{\rm p}A_{\rm p}(L) - i\mu_{\rm p}A_{\rm ext}$ where $r_{\rm p}$ and $\mu_{\rm p} = \sqrt{1 - r_{\rm p}^2}$ are the reflection and transmission coefficients for the pump in the ring resonator. In addition, the phase accumulated by the fields after a round trip will select a set of frequencies which are resonant for the ring. In the following, we assume that this resonance condition is satisfied. In this limit we find

$$\overline{A_{\rm s}B} \approx \bar{A}_{\rm s}\bar{B} \tag{18}$$

$$A_{\rm p}(L) - A_{\rm p}(0) \approx i\mu_{\rm p}A_{\rm ext} + (1 - r_{\rm p})A_{\rm p}(L). \tag{19}$$

Assuming that $\mu_{\mathrm{p}} \ll 1$ we find

$$A_{\mathrm{p}}(L) - A_{\mathrm{p}}(0) \approx i\mu_{\mathrm{p}}A_{\mathrm{ext}} - \ln r_{\mathrm{p}}\bar{A}_{\mathrm{p}}, \tag{20}$$

where we have assumed $A_{\mathrm{p}}(L) \approx \bar{A}_{\mathrm{p}}$ and we have approximated $r_{\mathrm{p}} - 1$ as $\ln r_{\mathrm{p}}$. We should mention that we represent $r_{\mathrm{p}} - 1$ as $\ln r_{\mathrm{p}}$ in order to make a clear connection between the effective optical decay rate in the mean field approach and the loaded decay rate in ring resonators. By carrying out a similar treatment for the Stokes and the phonon fields we find the mean-field equations of motion given by

$$\dot{\bar{A}}_{\mathrm{p}} + \frac{1}{2}v_{\mathrm{g,p}}(\alpha_{\mathrm{p}} - \frac{2}{L}\ln r_{\mathrm{p}})\bar{A}_{\mathrm{p}} = -i\tilde{g}\bar{A}_{\mathrm{s}}\bar{B} - i\frac{v_{\mathrm{g,p}}}{L}\mu_{\mathrm{p}}A_{\mathrm{ext}} + \frac{1}{\sqrt{L}}\bar{\eta}_{\mathrm{p}} \tag{21}$$

$$\dot{\bar{A}}_{\mathrm{s}} + \frac{1}{2}v_{\mathrm{g,s}}(\alpha_{\mathrm{s}} - \frac{2}{L}\ln r_{\mathrm{s}})\bar{A}_{\mathrm{s}} = -i\tilde{g}^*\bar{A}_{\mathrm{p}}\bar{B}^\dagger + \frac{1}{\sqrt{L}}\bar{\eta}_{\mathrm{s}} \tag{22}$$

$$\dot{\bar{B}} + (i\Omega_{\mathrm{b}} + \frac{1}{2}\Gamma)\bar{B} = -i\tilde{g}^*\bar{A}_{\mathrm{p}}\bar{A}_{\mathrm{s}}^\dagger + \frac{1}{\sqrt{L}}\bar{\xi}. \tag{23}$$

These equations can be written in a form that is reminiscent of normal mode Langevin equations through two steps. First note

$$\langle\bar{\xi}^\dagger(t)\bar{\xi}(t')\rangle = \frac{1}{L^2}\int_0^L dz \int_0^L dz' \, \langle\xi^\dagger(t,z)\xi(t',z')\rangle \tag{24}$$

$$= \frac{1}{L}\int_0^L dz \int_0^L dz' \, \Gamma n_{\mathrm{b}}^{\mathrm{th}}\delta(t-t')\delta(z-z') \tag{25}$$

$$= \Gamma n\delta(t-t'), \tag{26}$$

which has the same correlation properties as the Langevin forces of normal modes.

For simplicity of notation, we define $a_{\mathrm{p}} \equiv \sqrt{L}\bar{A}_{\mathrm{p}}$ (and similarly for the other fields). We also note that the loaded decay rates of the optical modes are given by $\gamma_{\mathrm{p}} = v_{\mathrm{g,p}}(\alpha_{\mathrm{p}} - \frac{2}{L}\ln r_{\mathrm{p}})$ and $\gamma_{\mathrm{s}} = v_{\mathrm{g,s}}(\alpha_{\mathrm{s}} - \frac{2}{L}\ln r_{\mathrm{s}})$. These notational simplifications lead to the coupled mean-field equations of motion, upon which we base our model of the temporal dynamics of laser oscillation. These equations are given by

$$\text{Mean-field equations of motion} = \begin{cases} \dot{a}_{\mathrm{p}} = -(i\omega_{\mathrm{p}} + \gamma_{\mathrm{p}}/2)\,a_{\mathrm{p}} - iga_{\mathrm{s}}b + \gamma_{\mathrm{p}}a_{\mathrm{p}}^{\mathrm{ext}}/2 + \eta_{\mathrm{p}} \\ \dot{a}_{\mathrm{s}} = -(i\omega_{\mathrm{s}} + \gamma_{\mathrm{s}}/2)\,a_{\mathrm{s}} - ig^*a_{\mathrm{p}}b^\dagger + \eta_{\mathrm{s}} \\ \dot{b} = -(i\Omega_{\mathrm{b}} + \Gamma/2)\,b - iga_{\mathrm{p}}a_{\mathrm{s}}^\dagger + \xi. \end{cases} \tag{27}$$

Here, $g = \tilde{g}/\sqrt{L}$ and $a_{\mathrm{p}}^{\mathrm{ext}}$ represents an external drive sourcing the pump field. In the absence of Brillouin coupling, this external drive determines the steady-state intracavity pump power. In addition, the Langevin forces $\eta_{\mathrm{p}}$, $\eta_{\mathrm{s}}$, and $\xi$ that account for thermal and quantum fluctuations of each of the fields. We assume that these forces are zero-mean Gaussian random variables with white power spectra, yielding correlation properties given by

$$\langle\eta_{\mathrm{p}}^\dagger(t)\eta_{\mathrm{p}}(t')\rangle = \gamma_{\mathrm{p}}n_{\mathrm{p}}^{\mathrm{th}}\delta(t-t') \tag{28}$$

$$\langle\eta_{\mathrm{s}}^\dagger(t)\eta_{\mathrm{s}}(t')\rangle = \gamma_{\mathrm{s}}n_{\mathrm{s}}^{\mathrm{th}}\delta(t-t') \tag{29}$$

$$\langle\xi^\dagger(t)\xi(t')\rangle = \Gamma n_{\mathrm{b}}^{\mathrm{th}}\delta(t-t'). \tag{30}$$

The magnitude of these power spectra are set by the thermal occupation numbers $n_p^{th}$, $n_s^{th}$, and $n_b^{th}$ (given by the Bose distribution).

### 1.2.1 Validity of mean-field analysis

A few comments are in order about the implications of this approximation. This analysis shows that the validity of our model requires that the optical field envelopes change slowly over the length of the ring resonator, which is well-satisfied for the Stokes field in our system which has decay length bigger than the size of the resonator. However, this approximation is only marginally satisfied for the pump field which has a decay length just bigger than the ring circumference.

The validity of the mean field treatment of the phonon field is more nuanced. The decay length for the phonon field is 63 μm. Since many decay lengths fit within the racetrack (4.6 cm) it may seem more natural to model the phonons as a collection of oscillators, made up of the different segments of the resonator. Moreover, the real resonator is composed of two distinct segments that support phononic guidance. However, by examining the phonon dynamics from the point of view of a driven field, we notice that if the optical fields change slowly over the length of the resonator so does the photoelastic driving force. From Eq. 8, we see that above threshold the steady-state phonon field strength is given by the local Stokes-pump beat note, which varies slowly around the device. Hence, under this approximation, we model the response of the elastic field as an average field with distributed coupling.

## 1.3 Inter-modal Brillouin laser noise theory

Having established our laser model, we now calculate the steady-state laser power, theshhold, slope efficiency, and power spectrum of our inter-modal Brillouin laser. We model the laser physics by using the coupled equations of motion for the mean-field amplitudes of the pump, Stokes, and phonon fields given in Eq. 27. In contrast with conventional Brillouin lasers, the temporal optical dissipation rates are much greater than the acoustic dissipation rates in our system, and therefore the optical fields can be adiabatically eliminated.

When the pump decay rate is large compared to the decay rates for the Stokes and the phonon fields, as well as the decoherence rate of the external laser, the pump field can be approximated as

$$a_p \approx -i\frac{ga_sb}{i(\Omega_b - \Omega) + \gamma_p/2} + a_p^{ext} \equiv -ig\chi_p(\Omega)a_sb + a_p^{ext}, \tag{31}$$

where $\Omega = \omega_p - \omega_s$ and $\chi_p(\Omega)$ is the pump susceptibility due to stimulated Brillouin scattering. This solution for the pump can be plugged into the remaining equations for the Stokes and the phonon fields yielding

$$\dot{a}_s = -(i\omega_s + \gamma_s/2)a_s - |g|^2\chi_p(\Omega)|b|^2a_s + \eta_s - ig^*a_p^{ext}b^\dagger \tag{32}$$

$$\dot{b} = -(i\Omega_b + \Gamma/2)b - |g|^2\chi_p(\Omega)|a_s|^2b + \xi - ig^*a_p^{ext}a_s^\dagger. \tag{33}$$

Next, we formally solve the equation for the Stokes field

$$a_s(t) = \int_{-\infty}^t d\tau\, e^{-[(i\omega_s + \frac{1}{2}\gamma_s)(t-\tau) + |g|^2\chi_p(\Omega)\int_\tau^t dt'\ |b(t')|^2]}\left[\eta_s(\tau) - ig^*a_p^{ext}(\tau)b^\dagger(\tau)\right]$$

$$= \hat{a}_s(t) - ig^*\int_{-\infty}^t d\tau\, e^{-[(i\omega_s + \frac{1}{2}\gamma_s)(t-\tau) + |g|^2\chi_p(\Omega)\int_\tau^t dt'\ |b(t')|^2]}a_p^{ext}(\tau)b^\dagger(\tau), \tag{34}$$

where $\hat{a}_{\mathrm{s}}$ represents the "dressed" annihilation operator of the Stokes field given explicitly by

$$\hat{a}_{\mathrm{s}}(t) = \int_{-\infty}^{t} d\tau \; e^{-[(i\omega_{\mathrm{s}}+\frac{1}{2}\gamma_{\mathrm{s}})(t-\tau)+|g|^2\chi_{\mathrm{P}}(\Omega)\int_{\tau}^{t} dt' \; |b(t')|^2]} \; \eta_{\mathrm{s}}(\tau). \tag{35}$$

The correlation properties of $\hat{a}_{\mathrm{s}}$ are are implicitly given in Eq. 50. We assume that the pump phase noise and phonon amplitude vary slowly in time compared to the decay rate of the Stokes field. This separation of time scales allows $\int_{\tau}^{t} dt' \; |b(t')|^2 \approx |b(t)|^2(t-\tau)$ in the exponent and the amplitudes of the external pump and the phonon to be taken outside of the integral (i.e. by adopting the Markov approximation where $b(\tau) \approx b(t)e^{i\Omega_{\mathrm{b}}(t-\tau)}$ under the integral). These approximations give

$$a_{\mathrm{s}}(t) \approx \hat{a}_{\mathrm{s}}(t) - ig^* a_{\mathrm{p}}^{\mathrm{ext}}(t)b^{\dagger}(t) \int_{-\infty}^{t} d\tau \; e^{-[i(\omega_{\mathrm{s}}-\omega_{\mathrm{P}}+\Omega_{\mathrm{b}})+\frac{1}{2}\gamma_{\mathrm{s}}+|g|^2\chi_{\mathrm{P}}(\Omega)|b(t)|^2](t-\tau)}$$

$$\approx \hat{a}_{\mathrm{s}}(t) - \frac{ig^* a_{\mathrm{p}}^{\mathrm{ext}}(t)b^{\dagger}(t)}{i(\Omega_{\mathrm{b}}-\Omega)+\frac{1}{2}\gamma_{\mathrm{s}}+|g|^2\chi_{\mathrm{P}}(\Omega)|b(t)|^2} \equiv \hat{a}_{\mathrm{s}}(t) - ig^*\chi_{\mathrm{s}}(\Omega)a_{\mathrm{p}}^{\mathrm{ext}}(t)b^{\dagger}(t). \tag{36}$$

Having derived the dynamics of the Stokes field for our system let us momentarily pause to point out some critical physics for our measurements. The heterodyne technique employed in our experiment directly measures the beat note of the Stokes field and the external laser, which acts as a local oscillator. Ignoring the intrinsic fluctuations of the Stokes field this beat note takes the form

$$a_{\mathrm{ext}}(t)a_{\mathrm{s}}^{\dagger}(t) = ig\chi_{\mathrm{s}}^{*}(\Omega)|a_{\mathrm{p}}^{\mathrm{ext}}(t)|^2 b(t). \tag{37}$$

There are few important points to make here. First, our external source laser is phase-noise dominated. Therefore, the beat note (a) is unaffected by the external laser noise (i.e. the beat note depends on the modulus square of the $a_{\mathrm{ext}}$), and (b) the beat note provides a direct window on the phonon dynamics. In the following we will discuss the dynamics of the phonon field.

By plugging the approximate solution for the Stokes field into Eq. 33 we obtain the effective nonlinear dynamics of the phonon field given by

$$\dot{b} = -\left(i\Omega_{\mathrm{b}}+\Gamma/2\right)b + |g|^2|\chi_{\mathrm{s}}(\Omega)|^2(i(\Omega_{\mathrm{b}}-\Omega)+\gamma_{\mathrm{s}}/2)|a_{\mathrm{p}}^{\mathrm{ext}}|^2 b + \xi - ig^* a_{\mathrm{p}}^{\mathrm{ext}}\hat{a}_{\mathrm{s}}^{\dagger}, \tag{38}$$

where we have used the definition of $\chi_{\mathrm{s}}$. In the following we consider the resonant case ($\Omega = \Omega_{\mathrm{b}}$, $\chi_{\mathrm{P}}(\Omega_{\mathrm{b}}) = 2/\gamma_{\mathrm{p}}$) where the equation of motion for the phonon drastically simplifies to

$$\dot{b} = -\left(i\Omega_{\mathrm{b}}+\Gamma/2\right)b + \frac{\gamma_{\mathrm{s}}}{2} \frac{|g|^2|a_{\mathrm{p}}^{\mathrm{ext}}|^2}{(\frac{1}{2}\gamma_{\mathrm{s}}+\frac{2}{\gamma_{\mathrm{p}}}|g|^2|b|^2)^2} b + \zeta, \tag{39}$$

where $\zeta \equiv \xi - ig^* a_{\mathrm{p}}^{\mathrm{ext}}\hat{a}_{\mathrm{s}}^{\dagger}$ is the effective Langevin forcing for the phonon, including pump noise and quantum fluctuations of the Stokes field.

### 1.3.1 Threshold behavior, lasing, and steady-state phonon amplitude

These nonlinear dynamics capture the basic the lasing behavior and noise properties of the 3-wave system. For large pump powers, when

$$\frac{2}{\gamma_{\mathrm{s}}}|g|^2|a_{\mathrm{p}}^{\mathrm{ext}}|^2 > \Gamma/2, \tag{40}$$

the mean-field phonon amplitude exhibits a dramatic increase in strength. From this analysis, we see that the threshold pump amplitude predicted by mean-field is

$$|a_{\mathrm{p,th}}^{\mathrm{ext}}|^2 \sim \frac{\Gamma \gamma_{\mathrm{s}}}{4|g|^2},\tag{41}$$

which, by converting to optical power $P = \hbar \omega v_g |a|^2 / L$, gives the threshold power

$$P_{\mathrm{th}} \sim \frac{\hbar \omega_{g,\mathrm{p}} v_{\mathrm{g,p}} \Gamma \gamma_{\mathrm{s}}}{4|g|^2 L} = \frac{\gamma_{\mathrm{s}}}{G_{\mathrm{b}} v_{\mathrm{g,s}}} = \frac{1}{G_{\mathrm{b}}}(\alpha_s - \frac{2}{L}\ln(r_{\mathrm{s}})).\tag{42}$$

From Eq. 42, we see that the mean-field prediction is consistent with the threshold pump power required for the round-trip optical SBS gain for the Stokes field to match the round-trip optical loss (See Eq. 12). We note here that because the phonon field mediates coupling between pump and Stokes wave, the coherent growth of the mean phonon field is simply a byproduct of the optical self-oscillation of the Stokes field. However, we will see in Section 1.4.b that due to the temporal dissipation hierarchy, the phonon field plays a dominant role in the linewidth narrowing dynamics.

The magnitude of the coherent amplitude of the phonon field above threshold can be obtained from the time average of Eq. 39. Assuming that $b = \beta e^{-i\Omega_{\mathrm{b}}t}$, where $\beta$ ($\gg \sqrt{n_{\mathrm{b}}^{\mathrm{th}}}$) is a time independent amplitude, multiplication of Eq. 39 by $e^{i\Omega_{\mathrm{b}}t}$ and taking a long-time average, where the Langevin force averages to zero, we find

$$0 = -\left[\Gamma/2 - \frac{\gamma_{\mathrm{s}}}{2}\frac{|g|^2|a_{\mathrm{p}}^{\mathrm{ext}}|^2}{(\frac{1}{2}\gamma_{\mathrm{s}} + \frac{2}{\gamma_{\mathrm{p}}}|g|^2|\beta|^2)^2}\right],\tag{43}$$

which relates the coherent phonon amplitude above threshold to the source power.

### 1.3.2   Slope efficiency

The steady-state phonon amplitude, above threshold, also allows us to estimate the slope efficiency for lasing on the Stokes mode. Assuming that the pump power is just above threshold we can assume $\beta < 1$. In this limit, the steady state amplitude equation Eq. 43 gives the relation $P_{\mathrm{ext}} - P_{\mathrm{th}} \approx 8|g|^2 P_{\mathrm{th}}\beta^2/(\gamma_{\mathrm{s}}\gamma_{\mathrm{p}})$. Plugging this relationship into the equation for the Stokes field, and converting amplitudes to power, we find the following relationship between the Stokes laser power $P_{\mathrm{s}}$ and the external source laser power $P_{\mathrm{ext}}$:

$$P_{\mathrm{s}} \approx \frac{\gamma_{\mathrm{p}}\omega_{\mathrm{s}} v_{\mathrm{g,s}}}{2\gamma_{\mathrm{s}}\omega_{\mathrm{p}} v_{\mathrm{g,p}}}(P_{\mathrm{ext}} - P_{\mathrm{th}}).\tag{44}$$

Here, the powers $P_{\mathrm{s}}$, $P_{\mathrm{ext}}$ and $P_{\mathrm{th}}$ represent the intracavity Stokes, pump, and threshold powers. Using the decay rates, frequencies, and group velocities of the optical modes obtained from transmission spectra in Sec. 3, Eq. 44 predicts an intracavity slope efficiency of $\approx 3\%$. By using the pump build up factor of 1.8, and the estimated ring-bus power coupling of 0.075 %, this analytical expression gives an on-chip slope efficiency estimate of $\approx 4\%$, close to the measured slope efficiency of 3% quoted in the main text.

In the following we consider the noise properties of the phonon field. These noise properties, described quantitatively by power spectra, depend sensitively on the nonlinear dynamics of the system and the source laser power. For this reason, we obtain the phonon power spectra near threshold numerically through stochastic simulations of the laser dynamics described by Eq. 27, and only at high source powers, well above threshold, will we present analytical results for the power spectra. We will show that our inter-modal laser exhibits power narrowing of the phonon linewidth.

## 1.4 Phonon power spectrum at high pump powers

### 1.4.1 Laser equations

Above threshold the system undergoes a lasing phase transition where the phonon and Stokes amplitude goes from having zero mean to having a finite expectation value given by $\beta$. Using this physics, we can find the power spectrum of the phonon field above threshold by reexpressing the phonon amplitude in terms of phase and amplitude

$$b = (\beta + \delta\beta)e^{-i\Omega_{\mathrm{b}}t + i\varphi}, \tag{45}$$

where $\beta$ is the time-independent mean value of the phonon amplitude, and $\delta\beta$ and $\varphi$ represent the zero-mean fluctuations of the phonon amplitude and phase, respectively. The dynamics of $\delta\beta$ and $\varphi$ can be found by plugging this representation of the phonon field into the effective equation of motion Eq. 39 and taking real and imaginary parts, giving

$$\dot{\delta\beta} \approx -\frac{1}{2}\Gamma_{\mathrm{RIN}}\delta\beta + \tilde{\zeta}_r \tag{46}$$

$$\dot{\varphi} \approx \frac{1}{\beta}\tilde{\zeta}_i, \tag{47}$$

where we have assumed $|\beta| \gg |\delta\beta|$, $\tilde{\zeta}_r(t) \equiv \mathrm{Re}[\zeta(t)\exp\{i\Omega_{\mathrm{b}}t - i\varphi(t)\}]$ and $\tilde{\zeta}_i(t) \equiv \mathrm{Im}[\zeta(t)\exp\{i\Omega_{\mathrm{b}}t - i\varphi(t)\}]$, and we have used the expression for the steady amplitude given in Eq. 42.

The decay rate for the amplitude fluctuations (or relative intensity fluctuations (RIN)) is given by

$$\Gamma_{\mathrm{RIN}} \equiv \frac{8}{\gamma_{\mathrm{p}}} \frac{\gamma_{\mathrm{s}}|g|^4|a_{\mathrm{p}}^{\mathrm{ext}}|^2\beta^2}{(\frac{1}{2}\gamma_{\mathrm{s}} + \frac{2}{\gamma_{\mathrm{p}}}|g|^2\beta^2)^3} \tag{48}$$

$$= \frac{8}{\gamma_{\mathrm{p}}} \frac{\Gamma|g|^2\beta^2}{(\frac{1}{2}\gamma_{\mathrm{s}} + \frac{2}{\gamma_{\mathrm{p}}}|g|^2\beta^2)}. \tag{49}$$

Next, we consider the correlation properties of the Langevin force $\tilde{\zeta}$.

### 1.4.2 Langevin force

In order to compute the power spectrum for the phonon field we need the correlation properties of $\tilde{\zeta}$. The relevant correlation function is given by

$$\begin{aligned}
\langle\tilde{\zeta}^{\dagger}(t)\tilde{\zeta}(t')\rangle &= \langle\zeta^{\dagger}(t)e^{-i\varphi(t)}e^{i\varphi(t')}\zeta(t')\rangle e^{-i\Omega_{\mathrm{b}}(t-t')} \\
&= \left[\langle\xi^{\dagger}(t)e^{-i\varphi(t)}e^{i\varphi(t')}\xi(t')\rangle + |g|^2\langle[a_{\mathrm{p}}^{\mathrm{ext}}(t)\hat{a}_{\mathrm{s}}^{\dagger}(t)]^{\dagger}e^{-i\varphi(t)}e^{i\varphi(t')}[a_{\mathrm{p}}^{\mathrm{ext}}(t')\hat{a}_{\mathrm{s}}^{\dagger}(t')]\rangle\right]e^{-i\Omega_{\mathrm{b}}(t-t')} \\
&\approx \Gamma n_{\mathrm{b}}^{\mathrm{th}}\delta(t-t') + |g|^2|a_{\mathrm{p}}^{\mathrm{ext}}|^2 e^{i(\omega_{\mathrm{p}}-\Omega_{\mathrm{b}})(t-t')}\langle\hat{a}_{\mathrm{s}}(t)\hat{a}_{\mathrm{s}}^{\dagger}(t')\rangle \\
&\approx \Gamma n_{\mathrm{b}}^{\mathrm{th}}\delta(t-t') + \frac{\gamma_{\mathrm{s}}|g|^2|a_{\mathrm{p}}^{\mathrm{ext}}|^2}{\gamma_{\mathrm{s}} + 4|g|^2\beta^2/\gamma_{\mathrm{p}}}(n_{\mathrm{s}}^{\mathrm{th}}+1)e^{-\frac{1}{2}(\gamma_{\mathrm{s}}+4|g|^2\beta^2/\gamma_{\mathrm{p}})|t-t'|} \\
&\approx \left[\Gamma n_{\mathrm{b}}^{\mathrm{th}} + 4\frac{\gamma_{\mathrm{s}}|g|^2|a_{\mathrm{p}}^{\mathrm{ext}}|^2}{(\gamma_{\mathrm{s}}+4|g|^2\beta^2/\gamma_{\mathrm{p}})^2}(n_{\mathrm{s}}^{\mathrm{th}}+1)\right]\delta(t-t') \\
&\approx \Gamma(n_{\mathrm{b}}^{\mathrm{th}}+n_{\mathrm{s}}^{\mathrm{th}}+1)\delta(t-t'). 
\end{aligned} \tag{50}$$

In the third line above we've assumed that the correlation time for the Stokes field is much shorter than the external pump and the phonon phase, allowing the phase noise of the external pump and phonon to be neglected. In the fourth line we have evaluated $\langle \hat{a}_s(t) \hat{a}_s^\dagger(t') \rangle$. In the fifth line we've assumed that the Stokes correlation function is effectively local in time, allowing us to replace the damped exponential with a delta function. This approximation will yield reliable results when computing the power spectrum for the phonon field as the decay rate for the Stokes field is much larger. Finally, in the last line we used to relation between the steady-state phonon amplitude and the phonon decay rate.

Using this resulting correlation function we find

$$\langle \tilde{\zeta}_r(t) \tilde{\zeta}_r(t') \rangle = \frac{1}{2} \Gamma (n_b^{th} + n_s^{th} + 1) \delta(t - t') \tag{51}$$

$$\langle \tilde{\zeta}_i(t) \tilde{\zeta}_i(t') \rangle = \frac{1}{2} \Gamma (n_b^{th} + n_s^{th} + 1) \delta(t - t') \tag{52}$$

$$\langle \tilde{\zeta}_r(t) \tilde{\zeta}_i(t') \rangle = 0. \tag{53}$$

### 1.4.3 Power spectra

Now we combine the correlation properties of the Langevin force with the phonon laser equations to derive the power spectrum for the phonon field. Using the correlation function

$$\langle e^{-i(\varphi(t) - \varphi(t'))} \rangle = e^{-\frac{1}{4\beta^2} \Gamma (n_b^{th} + n_s^{th} + 1) |t - t'|}, \tag{54}$$

obtained by solving Eq. 47 (*36*), we find

$$\begin{aligned}
\langle b^\dagger(t) b(t') \rangle &= [\beta^2 + \langle \delta\beta(t) \delta\beta(t') \rangle] \langle e^{-i(\varphi(t) - \varphi(t'))} \rangle e^{i\Omega_b(t - t')} \\
&= \left[ \beta^2 + \frac{\Gamma(n_b^{th} + n_s^{th} + 1)}{2\Gamma_{RIN}} e^{-\frac{1}{2}\Gamma_{RIN}|t - t'|} \right] e^{-\frac{1}{4\beta^2}\Gamma(n_b^{th} + n_s^{th} + 1)|t - t'|} e^{i\Omega_b(t - t')} \\
&\approx \left[ \underbrace{\beta^2 e^{-\frac{1}{4\beta^2}\Gamma(n + n_s + 1)|t - t'|}}_{\text{phase noise}} + \underbrace{\frac{\Gamma(n_b^{th} + n_s^{th} + 1)}{2\Gamma_{RIN}} e^{-\frac{1}{2}\Gamma_{RIN}|t - t'|}}_{\text{RIN}} \right] e^{i\Omega_b(t - t')}.
\end{aligned} \tag{55}$$

This result shows that at very high pump powers the phonon field becomes highly coherent. As $\beta \gg 1$ the power spectrum is dominated by the phase noise, which exhibits Schawlow-Townes narrowing. In this high-field limit, the coherence of the phonon oscillation is given by

$$\Delta\nu_b = \frac{\Gamma}{4\pi\beta^2} (n_b^{th} + n_s^{th} + 1), \tag{56}$$

which is the central result of this section. Thus, this mean-field model predicts phonon linewidth narrowing that is reminiscent of the phonon linewidth narrowing in cavity-optomechanical systems (*37*).

## 1.5 Simulations of laser dynamics

To understand the noise of our system near threshold, we explored the discrete time dynamics of the laser model described by Eq. 27. To perform these simulations we modify Eq. 27 in a number of ways. (A) We factored out the fast dynamics of the field amplitudes, i.e. we simulated the dynamics of the temporal envelopes of all fields, e.g. for the pump envelope $\bar{a}_p$ defined as $\bar{a}_p = a_p e^{i\omega_p t}$. (B) We considered the

laser dynamics on resonance, i.e. $\omega_\mathrm{p} = \omega_\mathrm{s} + \Omega_\mathrm{b}$. (C) We ignored quantum fluctuations of each of the fields, which play a minor role in our system. And, (D) we added the effects of nonlinear optical losses present in our silicon waveguide (these effects are described in detail in Ref. (*19*)).

Our simulations are performed in two steps. First, random noise fields are generated for the phonon field and the external source laser over the entire simulation domain, and then the fields are propagated in time using discrete time dynamics. Note that the intrinsic noise of the optical fields is quantum, so these sources of noise are neglected. In the discrete time picture the time axis is broken into time steps labeled $t_j$ and separated by $\Delta t$ (chosen to be 1 ns in our simulations). In this framework, the phonon Langevin force becomes $\xi(t_j) \rightarrow \xi_j$, where $\xi_j$ is a zero-mean Gaussian random variable, with variance $\Gamma n_\mathrm{b}^\mathrm{th}/\Delta t$, assigned to time step $j$. We assume a phase-noise dominated source laser, which we describe with the phase diffusion model (*25*). These noise fields are generated at the outset of the simulation. Subsequently, the fields are propagated from their initial values (set to 0) to find their time dynamics, at each time step the Langevin forces produce a random kick of the field amplitude and phase. For example, the phonon field is propagated by iterating the equation

$$b_{j+1} = b_j - \Delta t[(\Gamma/2)b_j + iga_{\mathrm{p},j}a_{\mathrm{s},j}^* - \xi_j] \tag{57}$$

which is the discrete time form for the mean-field phonon equation (Eq. 27), and similarly for the other fields. We obtain a wealth of information from these simulations, ranging from the power spectra and the occupation for all fields to the laser slope efficiency. In particular, the linewidth of the heterodyne beat note between the pump field and the Stokes field is displayed in Fig. 3 of the main text, and the simulated output laser power versus input source laser power is shown Fig. 2 of the main text.

### 1.5.1 Model parameters used in laser simulations

The model parameters used in these simulations were obtained or corroborated through independent measurements. For convenience, these parameters are presented in section 6.1. The optical decay rates are derived from transmission spectra of the ring resonator described in Sec. 3, and the Brillouin coupling, phonon frequency, and phonon decay rate were obtained from Brillouin scattering measurements in linear waveguides (*19*).

*Brillouin coupling rate:* Using the coupled envelope equations given in Eqs. 13-15, the Brillouin gain $G_\mathrm{b}$ can be directly related to the coupling parameter $g$, yielding $G_\mathrm{b} = 4|g|^2 L/(\hbar\omega_\mathrm{p} v_\mathrm{g,p} v_\mathrm{g,s}\Gamma)$ (*34*). The Brillouin gain of 470 $(\mathrm{Wm})^{-1}$, measured in linear inter-modal waveguides with the same cross section, correspond to a coupling rate of $g = 12$ kHz. In addition, the measured intracavity threshold power can be used as a cross check of this estimate. By using Eq. 42 with the measured laser threshold power of $P_\mathrm{th} = 19.3$ mW, along with the measured optical and acoustic dissipation rates, we find a Brillouin coupling rate of $g = 10.6$ kHz. Nonlinear optical losses are present in our system and will raise the threshold power. If we account for nonlinear optical losses the Stokes decay rate is slightly larger $2\pi \times 90$ MHz near threshold, yielding a slightly larger estimate for the Brillouin coupling rate of $g = 11.1$ kHz used in our simulations, this coupling rate corresponds with an estimated Brillouin gain of $G_\mathrm{b} = 400\mathrm{W}^{-1}\mathrm{m}^{-1}$.

*Phonon dissipation rate:* The phonon dissipation rate was obtained from spontaneous Brillouin scattering measurements in linear waveguides. These measurements give $\Gamma = 2\pi \times 13.1$ MHz.

*Linear optical loss:* We obtain the linear optical loss parameters by fitting the model presented in section 3 to the measured transmission spectra.

*Nonlinear optical loss:* The nonlinear optical parameters were studied in our previous work (*19*). From this detailed characterization of both intra- and inter-modal nonlinear loss, we observed that inter-modal scattering greatly reduces the spatial overlap between pump and Stokes waves (because they propagate in distinct optical spatial modes), resulting in low inter-modal TPA and FCA nonlinear loss

coefficients. These nonlinear losses have a relatively small effect on the threshold and slope efficiency at low Stokes powers.

## 1.6 Temporal dissipation hierarchy

This section presents a didactic figure (Fig. S1), which compares spectral properties produced by the temporal dynamics conventionally exhibited by Brillouin lasers (standard dissipation hierarchy) with those produced by this Brillouin laser (inverted dissipation hierarchy). Note that while the temporal dynamics are distinct, the spatial dynamics of this Brillouin laser (i.e., optical self-oscillation) are the same as conventional Brillouin lasers, as explained in section 1.7.

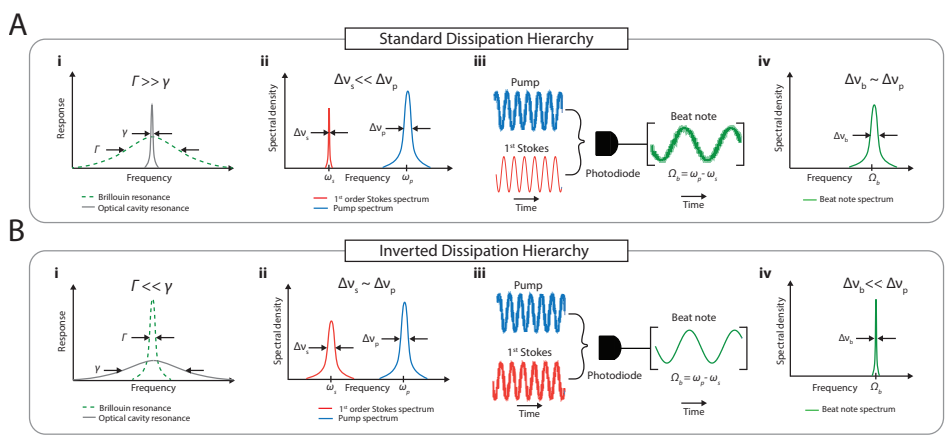

Figure S1: Spectral properties of distinct temporal dissipation hierarchies. (A) illustrates the noise properties of a Brillouin laser operating above threshold in the standard dissipation hierarchy where (i) the Brillouin gain bandwidth (or phononic decay rate, $\Gamma$) is much larger than the cavity linewidth ($\gamma$) of the optical resonator. In this limit, Brillouin lasing produces optical linewidth narrowing of the Stokes wave, resulting in the optical spectrum plotted in (ii), where the Stokes wave linewidth is narrow compared to that of the pump wave. (iii) diagrams a heterodyne measurement of the optical beat frequency between these two waves, which inherits the noise of the pump laser. This results in a noisy beat note with a linewidth approximately equal to the pump wave linewidth (iv). (B) shows the noise properties of this silicon Brillouin laser operating above threshold. (i) In this regime the dissipation hierarchy is inverted ($\Gamma \ll \gamma$). As a result, Brillouin lasing produces phonon linewidth narrowing, and the Stokes becomes a frequency-shifted copy of the pump, giving the optical spectrum plotted in (ii). As a result, heterodyne detection produces a clean beat note (iii) that directly reflects the phonon power spectrum (iv).

## 1.7 Three-wave laser dynamics

To explore the nature of laser oscillation in this system, we analyze the dynamics in the language of well-established optical parametric oscillator (OPO) physics (*31*), which provides a general framework for understanding and characterizing three-wave lasers. The intuition gained from optical parametric interactions is especially useful in the context of Brillouin scattering, since they have a number of characteristics in common, including spatial gain through traveling-wave interactions, dependence on phase-matching, and nearly identical spatio-temporal equations of motion.

In this section, we show that this framework allows us to (1) clearly distill the subtle yet fundamental differences between phonon lasers and Brillouin lasers in their many incarnations and (2) discover analogous systems that provide insight into the novel dynamics of this silicon Brillouin laser. In particular, we analyze our system in the context of singly-resonant optical parametric oscillators. These optical parametric oscillators exhibit coherent laser-oscillation similar to conventional lasers, with the crucial distinction that the optical gain is supplied by a three-wave (or $\chi^{(2)}$) interaction.

### 1.7.1 Laser characteristics

Before discussing the basic dynamics of OPO physics, we first highlight the essential properties of lasing. In its most basic form, a laser requires a cavity (or feedback) and a gain medium. Thus, lasing can occur for a field that (1) experiences feedback, which gives the system a pole, and (2) also experiences round-trip gain that matches round-trip loss, putting the pole on the real axis (*38*).

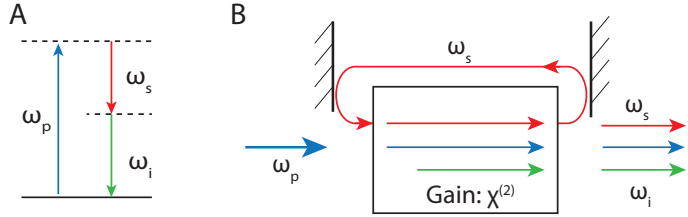

Figure S2: Basic operation of a singly resonant OPO (*31*). Panel (A) shows the energy conservation of this parametric process. The pump frequency ($\omega_p$) is equal to the sum of signal ($\omega_s$) and idler ($\omega_i$) frequencies. (B) In a singly-resonant OPO, only the signal wave experiences feedback from a cavity, and consequently self-oscillation. The idler wave is emitted as a byproduct of signal self-oscillation because it mediates the parametric process.

### 1.7.2 Singly-resonant OPO physics

A singly-resonant OPO consists of a nonlinear medium that provides parametric gain and a cavity that produces feedback for only one of the nonlinearly generated waves (*31*) (see Fig. S2). Within the nonlinear crystal, three waves interact: pump, the signal, and the idler waves. In this configuration, only the signal wave requires optical feedback for laser oscillation to occur. For instance, both the pump wave and idler can pass ballistically through the system without any form of feedback (see Fig. S2 ). Within a singly-resonant OPO, the condition for laser-oscillation is satisfied when the parametric gain equals the round-trip loss for the signal wave, resulting in optical self-oscillation of the signal wave (*31*). The idler is also emitted in tandem with the signal wave, as it mediates the parametric interaction; thus, the idler-wave emission is a byproduct of signal-wave self-oscillation.

### 1.7.3 Brillouin lasing and phonon lasing

We can now understand the basic operation of Brillouin lasing in terms of well-known OPO physics. In Brillouin lasers, high-frequency (10 GHz) phonons are heavily damped (spatially: mean-free path of $l_c \sim 100$ μm), preventing the formation of acoustic modes in systems with large dimensions ($L \gg l_c$). In contrast, the optical fields propagate with low spatial losses, permitting the formation of optical spatial modes within optical cavity resonators. Analogous to a singly-resonant OPO, the Stokes wave possesses a cavity while the acoustic field does not; thus, the Stokes wave performs the function of the signal wave

15

while the acoustic field plays the role of the idler field, resulting in Stokes self-oscillation. As the phonon field mediates coupling between pump and Stokes waves, phonon (idler) emission can be viewed as a byproduct of Stokes laser-oscillation. Thus, in general, Brillouin lasers only require an optical cavity for the Stokes wave; no phonon feedback is necessary for laser oscillation to occur.

This OPO framework also allows us to explore essential characteristics of optomechanical self-oscillation, also colloquially referred to as phonon lasing. In a phonon laser, the phonon field plays the role of the resonant signal field, while the Stokes wave performs the function of the idler. Thus, phonon lasing requires a phonon cavity (or acoustic feedback to produce a pole) and round-trip acoustic gain that can balance round-trip loss. In general, phonon lasers do not require optical feedback (for example, see the system proposed in Ref. (*39*)). In cavity-optomechanical systems, however, optical feedback is typically added to reduce the threshold of mechanical self-oscillation, (or phonon lasing). In this way, optomechanical self-oscillation shares many similarities with doubly resonant OPOs (where both signal (phonon) and idler (Stokes) waves possess a cavity (*31*)), but lacks the extended spatial characteristics (i.e. phase matching) typical of optical parametric oscillators.

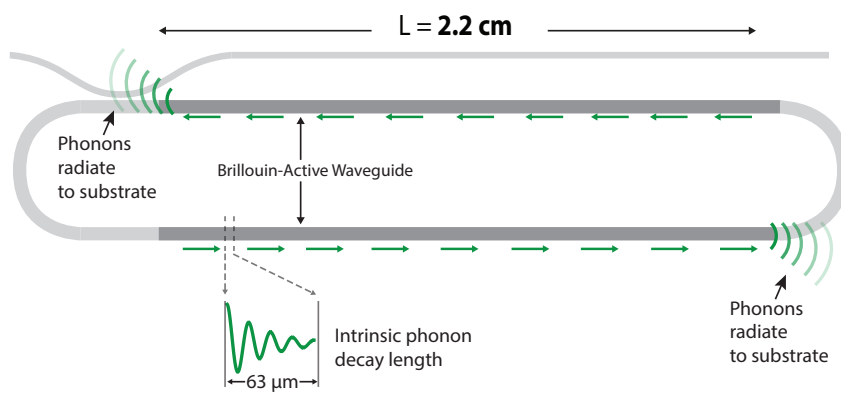

Figure S3: Diagram illustrating the spatial acoustic properties of this Brillouin laser. As with other Brillouin lasers, there is no acoustic feedback or phonon cavity in this system. The phonons are spatially heavily damped (i.e., intrinsic decay length $l_c \cong 63$ μm), preventing the formation of phonon cavity modes in this large racetrack resonator (circumference $\cong 4.6$ cm). Moreover, around the bends the waveguide is not suspended, prohibiting acoustic feedback. Instead, the phonon field efficiently radiates into the substrate.

### 1.7.4 Properties of the silicon Brillouin laser

Within the framework described above, we note that, despite the unusual temporal dynamics, this silicon laser shares all the same basic characteristics of Brillouin lasers. As diagrammed in Fig. 1 of the main text, optical feedback is produced by a long (4.6 cm) race-track resonator for the Stokes-wave. By comparison, the 6 GHz phonons that propagate in the Brillouin-active segments of this racetrack are heavily damped (with $l_c \cong 63$ μm and group velocity of 826 m/s), preventing the formation of discrete acoustic cavity modes. We note that, in addition to severe acoustic propagation losses, the device geometry prohibits any acoustic feedback. This is because, as shown in Fig. 1 of the main text, the acoustic phonons are not guided around the waveguide bends. As a result, the phonons are radiated to the substrate as seen in Fig. S3. Since only the optical Stokes wave can play the role of the signal wave, the phonon field must play the role of the idler wave. Therefore, this silicon laser produces optical self-oscillation, not phonon self-oscillation.

### 1.7.5 Optical self-oscillation and phonon linewidth narrowing

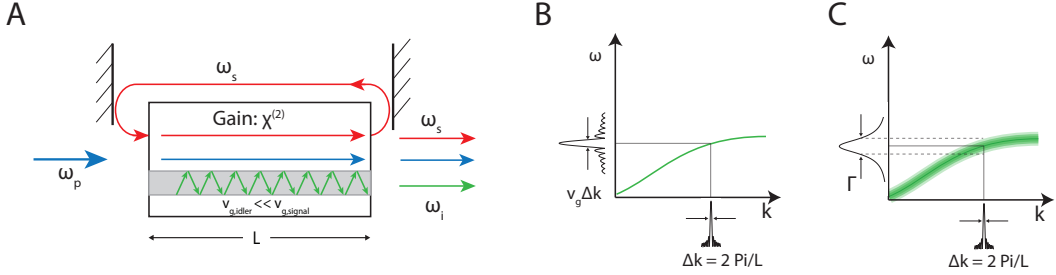

Figure S4: Generalized optical parametric oscillator with a slow group velocity idler field. (A) Diagram illustrating the basic device operation of a singly-resonant OPO with a slow idler field. Here, only the signal wave experiences feedback from a cavity. The idler field is emitted as a byproduct of signal-wave self-oscillation. (B) Dispersion relation of the idler wave that illustrates the phase-matching condition. The gain medium length and group velocity dictate the $\mathrm{sinc}^2$-like phase matching bandwidth. (C) Propagation losses effectively broaden the idler dispersion relation, resulting in a Lorentzian response with a width given by the dissipation rate ($\Gamma$).

While this system yields optical lasing, we have shown that it exhibits linewidth narrowing dynamics that more closely resemble the dynamics of a phonon laser. We again leverage the framework of OPO physics to illustrate how, even though the phonon field is not responsible for the feedback and laser oscillation, it governs the linewidth narrowing dynamics simply because it is the carrier with the longest temporal memory.

In what follows, we explore the dynamics of a generalized singly-resonant OPO that mirrors the behavior of this silicon Brillouin laser. As we have discussed, this and other Brillouin lasers are nearly direct analogues to a singly resonant OPO, in which the optical Stokes wave plays the role of the signal and the phonon field performs the function of the ballistic (i.e., cavityless) idler wave. However, this analogy becomes nuanced because a traveling acoustic wave, which has a velocity that is $10^5 \times$ slower than light, is taking the place of the optical idler wave. This distinction has profound consequences that permit the non-oscillating field to produce Schawlow-Townes linewidth narrowing.

To complete this analogy, let us consider a singly-resonant OPO in which the idler wave has an especially high group index, resulting in slow optical propagation. In this nonlinear optical system, phase matching plays a crucial role in the dynamics (see Fig. S4). As the velocities of the other waves are so much larger than that of the idler, the pump and signal waves essentially dictate the wavevector of the slow idler wave through phase matching. If this slow group velocity idler experiences no propagation losses, then the gain bandwidth (associated with phase-matched coupling) of this system is simply determined by the length of the gain medium and the group velocity of the idler wave (see Fig. S4B). Note that the smaller the group velocity the narrower the gain bandwidth.

However, any dissipation will effectively broaden the idler dispersion curve by the decay rate of the idler field. In other words, damping causes exponential decay; this produces Lorentzian broadening of the dispersion relation as seen in Fig. S4 (*39*). We can conceive a scenario in which the dissipation rate of the idler field is much larger than the dissipation-less phase matching bandwidth. In this heavily damped scenario, instead of the conventional $\mathrm{sinc}^2$-like phase matching bandwidth, the phase matching condition yields a Lorentzian gain bandwidth determined by the lifetime of the slow idler field (see Fig. S4C).

This system arrangement allows us to easily conceive of a condition in which the signal wave produces self-oscillation while the idler dictates the linewidth narrowing dynamics. As we have discussed, because there is no feedback for the idler wave, the idler wave cannot self-oscillate. Instead, a non-zero

expectation value of the idler can only be achieved as a byproduct of signal-wave self-oscillation. However, because of the disparate velocities of the signal and idler waves, it is apparent that even though the signal wave has a cavity and the idler does not, the idler wave can easily have longer temporal memory as it ballistically (yet ever so slowly) passes through the nonlinear medium. In this temporal dissipation hierarchy, we have seen that, compared to the idler, the signal wave is temporally overdamped, resulting in dynamics that adiabatically imitate those of the idler field. Thus, this OPO resonator produces laser-oscillation of the signal wave and linewidth narrowing of the non-resonant idler wave, in direct analogy to the silicon Brillouin laser studied in this work.

## 2 Sub-coherence self-heterodyne measurement

Above threshold, small variations in pump frequency translate to frequency instabilities that are not intrinsic to this system. As a result, the heterodyne spectrum must be acquired on a 30 millisecond timescale to avoid distortions produced by slow frequency hops generated by the pump laser. These shorter acquisition times effectively limit the resolution bandwidth of this measurement approach to 20 kHz. At higher emitted powers, we are able to overcome this resolution limitation by using a separate sub-coherence delayed self-heterodyne measurement technique. This method allows us to determine the intrinsic laser linewidth (and resulting phonon dynamics) despite the slow center-frequency hops induced by the pump laser (*40, 41*). This technique employs an unbalanced Mach-Zehnder interferometer that possesses both a sub-coherence delay arm and a frequency shifted reference arm (See Fig. S5). With sub-coherence delay times, the measured spectral density is composed of a delta function and a sinc-like background. By comparing the relative magnitude of each component, one can determine the coherence time of the actual laser. Using this sub-coherence technique, we measure the linewidth of the pump laser and show that the silicon Brillouin laser behaves analogously to a self-driven acousto-optic frequency shifter.

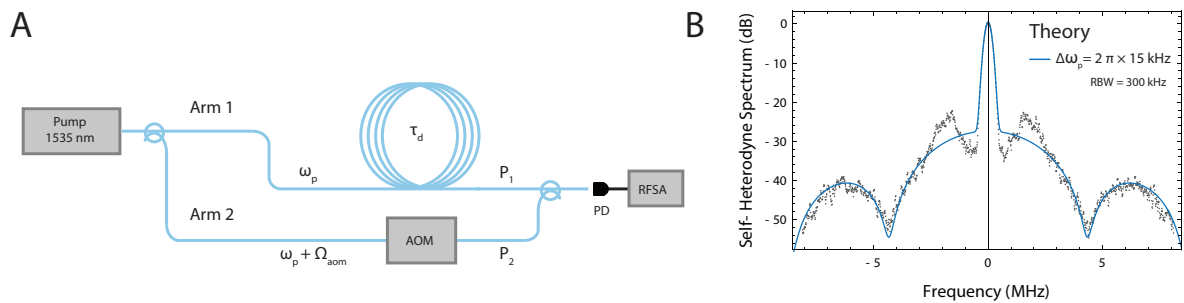

Figure S5: (A) Experimental setup for self-heterodyne interferometry. The first arm of the interferometer provides a delay while the second arm frequency shifts the light using an acousto-optic frequency shifter (AOM). The two waves are combined and the beat note is detected with a fast photodiode. (B) Self-heterodyne spectrum of the Agilent 81600B diode laser with a delay length of 48.25 m.

### 2.1 Self-heterodyne interferometry

In this section, we discuss the essential elements of the sub-coherence self heterodyne measurement. Figure S5A depicts a typical self-heterodyne interferometer. For an arbitrary delay length, the self-heterodyne spectral density is found to be (*40*)

$$S[\omega] \propto P_1 P_2 \left[ \pi e^{-\Delta\omega_{\mathrm{p}}\tau_{\mathrm{d}}} \delta(\omega - \Omega_{\mathrm{aom}}) + \frac{\Delta\omega_{\mathrm{p}}}{(\omega - \Omega_{\mathrm{aom}})^2 + \Delta\omega_{\mathrm{p}}^2} (1 - e^{-\Delta\omega_{\mathrm{p}}\tau_{\mathrm{d}}} (\cos((\omega - \Omega_{\mathrm{aom}})\tau_{\mathrm{d}}) \right.$$
$$\left. + \frac{\sin((\omega - \Omega_{\mathrm{aom}})\tau_{\mathrm{d}})\Delta\omega_{\mathrm{p}}}{\omega - \Omega_{\mathrm{aom}}})) \right].$$

(58)

Here, $P_1$ and $P_2$ are the powers in arm 1 and arm 2 of the self-heterodyne interferometer, $\Delta\omega_{\mathrm{p}}$ is the linewidth of the laser in units of angular frequency, $\Omega_{\mathrm{aom}}$ is the driving frequency of the acousto-optic modulator, and $\tau_{\mathrm{d}}$ is the transit time of the delay arm.

Note that in the limit that $\tau_{\mathrm{d}} \gg 1/\Delta\omega_{\mathrm{p}}$, the spectrum becomes a simple Lorentzian with a HWHM of the intrinsic laser linewidth. For this reason, the self-heterodyne technique typically involves a delay line that far exceeds the expected coherence time. However, as with other external cavity diode lasers, the low frequency fluctuations of the center laser frequency of the Agilent 81600B obscure the traditional measurement (*41*). For this reason, we employ a sub-coherence delay arm to measure the linewidth of the pump laser. Figure S5B shows a comparison of a typical self-heterodyne signal of the pump laser (Agilent 81600B) with the theoretical spectral density of a laser with a linewidth of 15 kHz.

## 2.2 Sub-coherence self-heterodyne measurement using a silicon Brillouin laser

### 2.2.1 Spectral characteristics of silicon Brillouin laser

In this section, we demonstrate the unique noise properties that result from the temporal dissipation of this SBS laser (i.e., $\Gamma \ll \gamma_{\mathrm{s}}, \gamma_{\mathrm{s}}$). This is done by using the Brillouin laser as a frequency-shifter and showing that it behaves in complete analogy to an acousto-optic frequency shifter. We perform this measurement using the experimental setup shown in Fig. S6.

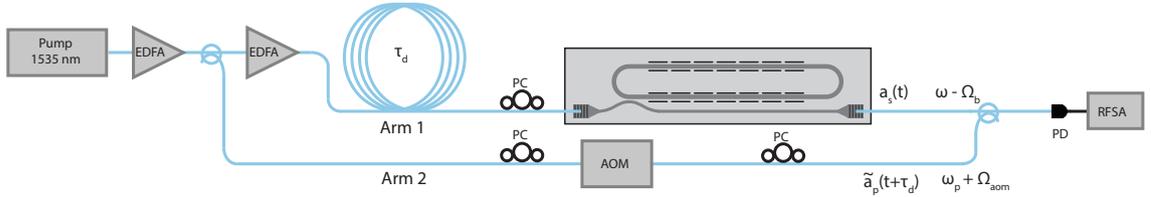

Figure S6: Self-heterodyne interferometer using the silicon Brillouin laser as the frequency-shifter. An auxiliary AOM is employed here only to spectrally discriminate the light passing through arm 2 from the pump light ballistically passing through the chip.

In the discussion to follow, we present a simple model that captures the effect of phonon noise in the self-heterodyne measurement. Using the Wiener-Khinchin theorem and assuming stationarity, the spectral density of a heterodyne beat note can be written as

$$S[\omega] = \frac{1}{2\pi} \int_{-\infty}^{\infty} d\tau \, e^{i\omega\tau} \langle E_{\mathrm{LO}}(\tau) E_{\mathrm{s}}^*(\tau) E_{\mathrm{LO}}^*(0) E_{\mathrm{s}}(0) \rangle,$$

(59)

where $E_s$ and $E_{LO}$ represent the signal and local oscillator waves, respectively. Applying Eq. (59) to the measurement presented in Fig. S6, the spectral density becomes

$$S[\omega] = \frac{1}{2\pi} \int_{-\infty}^{\infty} d\tau \; e^{i\omega\tau} \langle \tilde{a}_p(\tau + \tau_d) a_s^\dagger(\tau) \tilde{a}_p^\dagger(\tau_d) a_s(0) \rangle, \tag{60}$$

where $a_s$, $a_p$, and $\tilde{a}_p$ will denote the Stokes, input pump, and AOM-shifted pump waves, respectively. Here, $\tau_d$ represents the transit time of the delay arm. Through separation of time scales, the Stokes field can be written as

$$a_s = \frac{-2ig^* a_p b^\dagger}{\gamma_s(1 + \frac{4|g|^2|\beta|^2}{\gamma_s\gamma_p})}. \tag{61}$$

Hence, the spectral density becomes

$$S[\omega] = \frac{2|g|^2}{\pi(\gamma_s(1 + \frac{4|g|^2|\beta|^2}{\gamma_s\gamma_p}))^2} \int_{-\infty}^{\infty} d\tau \; e^{i\omega\tau} \langle \tilde{a}_p(\tau + \tau_d) a_p^\dagger(\tau) b(\tau) \tilde{a}_p^\dagger(\tau_d) a_p(0) b^\dagger(0) \rangle. \tag{62}$$

Due to the separation of time scales, the phonon field and pump fields are uncorrelated. Therefore the expectation value in Eq. 62 simplifies to

$$S[\omega] = \frac{2|g|^2}{\pi(\gamma_s(1 + \frac{4|g|^2|\beta|^2}{\gamma_s\gamma_p}))^2} \int_{-\infty}^{\infty} d\tau \; e^{i\omega\tau} \langle \tilde{a}_p(\tau + \tau_d) a_p^\dagger(\tau) \tilde{a}_p^\dagger(\tau_d) a_p(0) \rangle \langle b(\tau) b^\dagger(0) \rangle. \tag{63}$$

We will now focus on the first expectation value. We can write the phases of $a_p$ and $\tilde{a}_p$ as

$$\begin{aligned} \phi(t) &= -i\omega_p t + i\phi_n(t) \\ \tilde{\phi}(t) &= -i(\omega_p + \Omega_{aom})t + i\phi_n(t). \end{aligned} \tag{64}$$

Here, we assume $\phi_n$ obeys the phase diffusion equation such that

$$\begin{aligned} \dot{\phi}(t) &= \eta(t) \\ \langle \eta(t)\eta(t') \rangle &= \Delta\omega_p \delta(t - t'). \end{aligned} \tag{65}$$

Applying Eq. 70 to the first expectation value in Eq. 63, we find

$$\begin{aligned} \langle \tilde{a}_p(\tau + \tau_d) a_p^\dagger(\tau) \tilde{a}_p^\dagger(\tau_d) a_p(0) \rangle &= |a_p|^2 |\tilde{a}_p|^2 \langle e^{i\Delta\phi_n(\tau,\tau_d)} \rangle e^{-i\Omega_{aom}\tau} \\ \Delta\phi_n(\tau, \tau_d) &= \phi_n(\tau + \tau_d) - \phi_n(\tau) - \phi_n(\tau_d) - \phi_n(0). \end{aligned} \tag{66}$$

By direct integration, we compute $\Delta\phi_n(\tau, \tau_d)$:

$$\Delta\phi_n(\tau, \tau_d) = \int_{\tau}^{\tau + \tau_d} dt' \; \eta(t') - \int_0^{\tau_d} dt' \; \eta(t'). \tag{67}$$

As $\Delta\phi_n(\tau, \tau_d)$ can be described by Gaussian statistics, we can rewrite the expectation value for $\langle e^{i\Delta\phi_n(\tau,\tau_d)} \rangle$ (36), yielding

$$\langle e^{i\Delta\phi_n(\tau,\tau_d)} \rangle = e^{\frac{-\langle [\Delta\phi_n(\tau,\tau_d)]^2 \rangle}{2}}. \tag{68}$$

Evaluating Eq. 68, the spectral density becomes

$$S[\omega] = \frac{2|g|^2|a_\mathrm{p}|^2|\tilde{a}_\mathrm{p}|^2}{\pi(\gamma_\mathrm{s}(1+\frac{4|g|^2|b|^2}{\gamma_\mathrm{s}\gamma_\mathrm{p}}))^2} \int_{-\infty}^{\infty} d\tau e^{i(\omega-\Omega_\mathrm{aom})\tau} e^{-\Delta\omega_\mathrm{p}(\tau_\mathrm{d}-(\tau_\mathrm{d}-|\tau|)\theta(\tau_\mathrm{d}-|\tau|))} \langle b(\tau)b^\dagger(0)\rangle. \tag{69}$$

Now we use the same process (Eqs. 70-68) to determine the phonon correlation function.

$$\begin{aligned}
\Phi(t) &= -i\Omega_\mathrm{b}t + i\phi_\mathrm{n}(t) \\
\dot{\Phi}(t) &= \eta_b(t) \\
\langle \eta_b(t)\eta_b(t')\rangle &= \Delta\Omega_\mathrm{b}\delta(t-t') \\
\langle b(\tau)b^\dagger(0)\rangle &= |\beta|^2\langle e^{i\Delta\Phi(\tau)}\rangle e^{-i\Omega_\mathrm{b}\tau} \\
\Delta\Phi(\tau) &= \int_0^\tau dt'\eta_b(t') \\
\langle b(\tau)b^\dagger(0)\rangle &= \beta^2 e^{-i\Omega_\mathrm{b}\tau - \frac{\Delta\Omega_\mathrm{b}|\tau|}{2}}
\end{aligned} \tag{70}$$

Here, $\Omega_\mathrm{b}$ is the phonon center frequency and $\Delta\Omega_\mathrm{b}$ is phonon linewidth (units of angular frequency, $\Delta\Omega_\mathrm{b} = 2\pi\Delta\nu_\mathrm{b}$). Inserting the phonon correlation function into Eq. 69 yields

$$S[\omega] = \frac{2|g|^2|a_\mathrm{p}|^2|\tilde{a}_\mathrm{p}|^2|\beta|^2}{\pi(\gamma_\mathrm{s}(1+\frac{4|g|^2|\beta|^2}{\gamma_\mathrm{s}\gamma_\mathrm{p}}))^2} \int_{-\infty}^{\infty} d\tau e^{i(\omega-\Omega_\mathrm{aom}-\Omega_\mathrm{b})\tau} e^{-\Delta\omega_\mathrm{p}(\tau_\mathrm{d}-(\tau_\mathrm{d}-|\tau|)\theta(\tau_\mathrm{d}-|\tau|))-\frac{\Delta\Omega_\mathrm{b}|\tau|}{2}}. \tag{71}$$

Note that in the limit that $\tau_\mathrm{d} \to 0$, heterodyne detection directly measures the phonon power spectrum. Thus, the spectral density becomes

$$S[\omega] = \frac{2|g|^2|a_\mathrm{p}|^2|\tilde{a}_\mathrm{p}|^2|\beta|^2}{\pi(\gamma_\mathrm{s}(1+\frac{4|g|^2|\beta|^2}{\gamma_\mathrm{s}\gamma_\mathrm{p}}))^2} \int_{-\infty}^{\infty} d\tau e^{i(\omega-\Omega_\mathrm{aom}-\Omega_b)\tau} e^{-\frac{\Delta\Omega_\mathrm{b}|\tau|}{2}}. \tag{72}$$

Performing the integration of Eq. 71 with an arbitrary delay length, we arrive at the main result for this section:

$$\begin{aligned}
S[\omega] =& \frac{2|g|^2|a_\mathrm{p}|^2|\tilde{a}_\mathrm{p}|^2|\beta|^2}{\pi(\gamma_\mathrm{s}(1+\frac{4|g|^2|\beta|^2}{\gamma_\mathrm{s}\gamma_\mathrm{p}}))^2} \left[ \frac{4(\Delta\Omega_\mathrm{b}+2\Delta\omega_\mathrm{p})}{(\Delta\Omega_\mathrm{b}+2\Delta\omega_\mathrm{p})^2 + 4w^2} \right. \\
& - \frac{32w\Delta\omega_\mathrm{p}(\Delta\Omega_\mathrm{b}+\Delta\omega_\mathrm{p})\sin(w\tau_\mathrm{d})e^{-\frac{1}{2}\tau_\mathrm{d}(\Delta\omega_b+2\Delta\omega_\mathrm{p})}}{(\Delta\Omega_\mathrm{b}^2+4w^2)((\Delta\Omega_\mathrm{b}+2\Delta\omega_\mathrm{p})^2+4w^2)} \\
& \left. + \frac{8\Delta\omega_\mathrm{p}(\Delta\Omega_\mathrm{b}^2+2\Delta\Omega_\mathrm{b}\Delta\omega_\mathrm{p}-4w^2)\cos(w\tau_\mathrm{d})e^{-\frac{1}{2}\tau_\mathrm{d}(\Delta\Omega_\mathrm{b}+2\Delta\omega_\mathrm{p})}}{(\Delta\Omega_\mathrm{b}^2+4w^2)((\Delta\Omega_\mathrm{b}+2\Delta\omega_\mathrm{p})^2+4w^2)} \right].
\end{aligned} \tag{73}$$

Here, $w \equiv \omega - \Omega_\mathrm{aom} - \Omega_\mathrm{b}$. If we instead evaluate Eq. (71) with $\Delta\Omega_\mathrm{b} = 0$ (zero phonon phase noise), the spectrum matches that of the simple sub-coherence self-heterodyne measurement (refer to Eq. (58)). Performing the integration we find

$$S[\omega] = \frac{2|g|^2|a_{\rm p}|^2|\tilde{a}_{\rm p}|^2|\beta|^2}{\pi(\gamma_{\rm s}(1+\frac{4|g|^2|\beta|^2}{\gamma_{\rm s}\gamma_{\rm p}}))^2}\left[2\pi e^{-\Delta\omega_{\rm p}\tau_{\rm d}}\delta(w) + \frac{2\Delta\omega_{\rm p}}{w^2+\Delta\omega_{\rm p}^2}(1-e^{-\Delta\omega_{\rm p}\tau_{\rm d}}(\cos(w\tau_{\rm d})+\frac{\sin(w\tau_{\rm d})\Delta\omega_{\rm p}}{w}))\right]$$

$$\propto P_1 P_2\left[\pi e^{-\Delta\omega_{\rm p}\tau_{\rm d}}\delta(w) + \frac{\Delta\omega_{\rm p}}{w^2+\Delta\omega_{\rm p}^2}(1-e^{-\Delta\omega_{\rm p}\tau_{\rm d}}(\cos(w\tau_{\rm d})+\frac{\sin(w\tau_{\rm d})\Delta\omega_{\rm p}}{w}))\right].$$

(74)

Here $P_1$ and $P_2$ are the powers in arm 1 and arm 2 of the self-heterodyne interferometer. Figure S7A plots Eq. 73 with a 15 kHz linewidth pump and various phonon linewidths. Note that as the phonon linewidth increases, the interference fringes wash out and the pedestal becomes more Lorentzian-like.

Using the experimental setup depicted in Fig. S6, we measure the self-heterodyne spectrum with the silicon Brillouin laser as the frequency-shifter and compare it to that of traditional measurement using an acousto optic modulator. Fig. S7B shows the two spectra superimposed. Strong qualitative agreement between the two experimental spectra demonstrates that the SBS laser behaves analogously to a self-driven acousto-optic frequency shifter. More specifically, the emitted Stokes light from this silicon Brillouin laser is a frequency-shifted copy of the pump light with a small degree of phase noise imparted by the phonon field. Fig. S7C shows the range of the spectrum that is most sensitive to the phonon linewidth and the theoretical comparison. The spectrum plotted (Stokes power of 0.025 mW) in Fig. S7C shows good agreement with the theoretical spectrum with a phonon linewidth of 3 kHz.

We also perform these measurements using an open-loop laser that exhibits exceptional agreement with the sub-coherence spectrum. For these measurements, see Section 5.3.

### 2.2.2   Spectral characteristics conventionally exhibited by Brillouin lasers

The self-heterodyne measurement performed with the silicon Brillouin laser demonstrates the unique phonon linewidth narrowing characteristics of this device. By contrast, in the dissipation hierarchy conventionally realized in silica Brillouin lasers— that is, $\Gamma \gg \gamma_{\rm p}, \gamma_{\rm s}$ — the predicted spectrum is very different from both Eq. 73 and the experimentally measured spectrum of this silicon Brillouin laser. In this more widely studied regime, the Stokes field experiences Schawlow-Townes narrowing, resulting in a narrow linewidth Stokes wave whose phase is now uncorrelated with that of the pump. Thus, in the conventional limit (i.e., $\Gamma \gg \gamma_{\rm p}, \gamma_{\rm s}$), Eq. 60 becomes

$$S[\omega] = \frac{1}{2\pi}\int_{-\infty}^{\infty}d\tau e^{i\omega\tau}\langle\tilde{a}_{\rm p}(\tau+\tau_{\rm d})\tilde{a}_{\rm p}^{\dagger}(\tau_{\rm d})\rangle\langle a_{\rm s}^{\dagger}(\tau)a_{\rm s}(0)\rangle.$$

(75)

Applying the same technique as that of the previous section, the spectral density becomes

$$S[\omega] = \frac{|\tilde{a}_{\rm p}|^2|a_{\rm s}|^2}{2\pi}\int_{-\infty}^{\infty}d\tau e^{i(\omega-\Omega_{\rm b}-\Omega_{\rm aom})\tau}e^{\frac{-(\Delta\omega_{\rm p}+\Delta\omega_{\rm s})|\tau|}{2}}.$$

(76)

Here, $\Delta\omega_{\rm p}$ and $\Delta\omega_{\rm s}$ are the linewidths for the pump and Stokes waves, respectively (angular frequency units). Evaluating Eq. 76, the spectral density in the conventional Brillouin limit is a simple Lorentzian, completely devoid of coherent interference. Thus, the spectrum becomes

$$S[\omega] \propto P_{\rm s}P_{\rm p}\frac{4(\Delta\omega_{\rm p}+\Delta\omega_{\rm s})}{(\Delta\omega_{\rm p}+\Delta\omega_{\rm s})^2+4(\omega-\Omega_{\rm b}-\Omega_{\rm aom})^2}.$$

(77)

In Brillouin lasers studied to date, $\Delta\omega_{\rm s} \ll \Delta\omega_{\rm p}$, and the observed Lorentzian has full width at half maximum (FWHM) of $\Delta\omega_{\rm p}$. Note that Eq. 77 is independent of $\tau_{\rm d}$, as there is no phase coherence

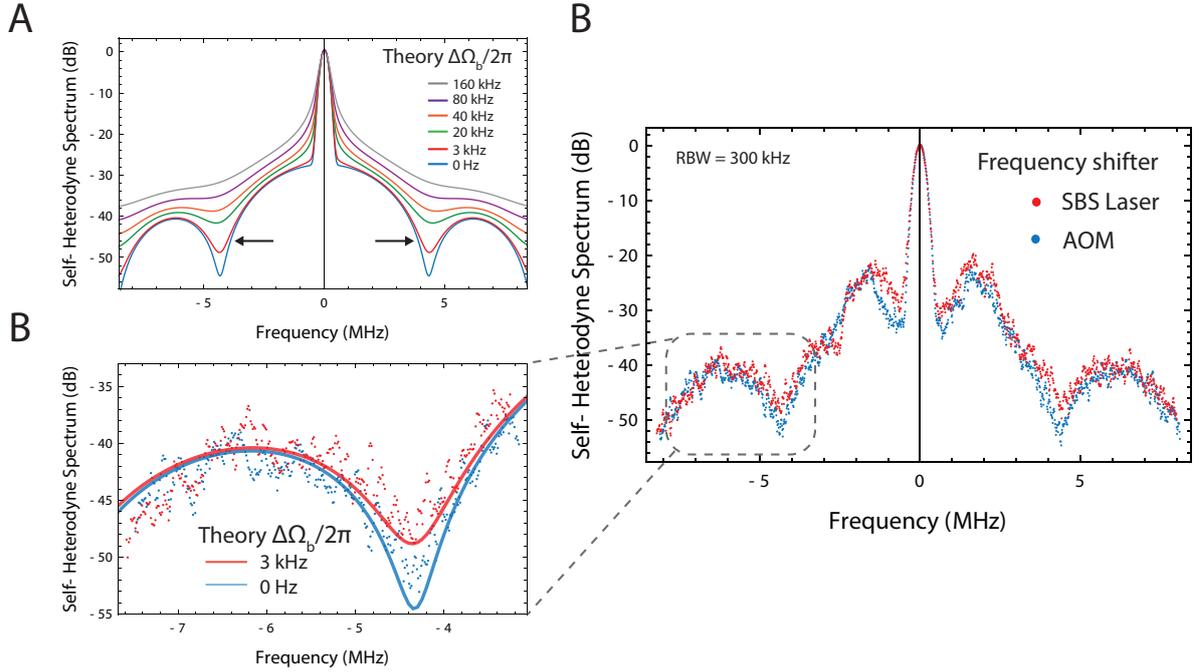

Figure S7: (A) Theoretical trends showing the effect of phonon phase noise on the self-heterodyne measurement. Observe that as the phonon linewidth increases, the fringes lose contrast. (B) Comparison of the self-heterodyne spectrum using the Brillouin laser as the frequency-shifter (Figure S6) with the traditional spectrum using an AOM as the frequency-shifter (Figure S5). Identical delay lengths are used in each experiment. (C) Theory-experiment comparison in the most sensitive portion of the spectrum to the phonon linewidth. Note that there is some residual phase noise that reduces the contrast of the fringes from the 0 Hz theoretical trend (evident from the spectra using the AOM as the frequency shifter). We attribute this noise to small variations in the delay arm of our fiber interferometer.

for interference. Hence, in the traditional Brillouin limit, the self-heterodyne spectrum is bereft of the characteristic fringes, in sharp contrast to the phonon linewidth narrowing regime explored by the silicon Brillouin laser.

# 3   Six-port coupler theory and analysis

In this section we derive the optical response of a multimode optical resonator that we use to determine the input parameters of our laser model. By fitting this theory to measured transmission spectra we obtain a wealth of information about the ring resonator physics. We determine the free-spectral range, giving the group velocity, and the width of the transmission dips gives the loaded dissipation rates for each of the optical modes. Once these parameters are specified this theory can be used to determine the intracavity pump and Stokes powers from the transmitted pump and the output Stokes powers, and provides all of the input parameters for our lumped element model.

We undertake this analysis by using general properties of lossless six-port couplers of the type pictured in Fig. S8, and by leveraging the known free propagation of the optical fields in each waveguide type.

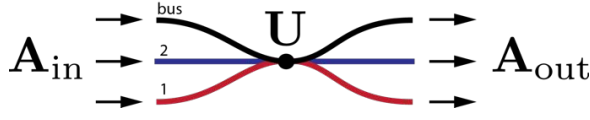

Figure S8: Illustration of a six-port coupler. In general, the propagation velocity and dissipation is distinct in each waveguide.

## 3.1 Six-port coupler

To begin, consider the lossless six-port coupler seen in Fig. S8, which can be described by a symmetric $3 \times 3$ unitary matrix $\mathbf{U}$. In such a coupler, light injected into any one input port can be scattered out of all output ports; the relative phases of all output ports are in general arbitrary, but the total power is conserved through the coupler. Symmetry of the scattering matrix ensures that the coupler is reciprocal. Mathematically, this scattering is described by the relation

$$\mathbf{A}_{\text{out}} = \mathbf{U} \cdot \mathbf{A}_{\text{in}}, \tag{78}$$

where $\mathbf{A}(z) = (A_{\text{bus}}(z), A_2(z), A_1(z))$ is a vector composed of the field amplitudes in the respective bus, first excited, and fundamental modes at position $z$ along the composite 'waveguide' (i.e. the three waveguides). $\mathbf{A}_{\text{in}}$ being the value of this vector just before entering the coupler, and $\mathbf{A}_{\text{out}}$ representing the resultant field amplitudes exiting the coupler. By virtue of unitarity (i.e. $\mathbf{U}^\dagger \cdot \mathbf{U} = \mathbf{1}$) the total photon flux, or the total optical power, is conserved through this coupler, i.e. $|\mathbf{A}_{\text{out}}|^2 = |\mathbf{U} \cdot \mathbf{A}_{\text{in}}|^2 = |\mathbf{A}_{\text{in}}|^2$.

## 3.2 Free optical propagation

To describe the physics of the ring resonator we need to understand the free propagation of the fields in each waveguide. We assume that the dynamics of optical field is distinct to each waveguide, and can be described by a spatial damping factor and a phase. Hence, as the optical field propagates a distance $\ell$ the field transforms as

$$\mathbf{A}(\ell + z) = \mathbf{T}(\ell) \cdot \mathbf{A}(z), \tag{79}$$

where the propagation matrix is given by

$$\mathbf{T}(\ell) = \begin{pmatrix} e^{(ik_{\text{bus}} - \alpha_{\text{bus}}/2)\ell} & 0 & 0 \\ 0 & e^{(ik_2 - \alpha_2/2)\ell} & 0 \\ 0 & 0 & e^{(ik_1 - \alpha_1/2)\ell} \end{pmatrix} \tag{80}$$

Here, $k_j = \omega / v_j$ is the propagation constant; $v_j$ being the phase velocity in the $j$th waveguide, and $\omega$ being the angular frequency. The factor $\alpha_j$ is the spatial decay rate unique to each waveguide.

In the next section we put these ingredients together to find the transmission spectrum of a ring resonator formed by closing waveguides 1 and 2 into closed loops of length $L$.

## 3.3 Optical response of a multimode ring resonator

Consider a ring resonator formed by closing waveguide 1 and 2 into closed loops of length $L$ (see Fig. S9). These two closed loops will each form a distinct set of optical modes that can be characterized by observing the transmission of the optical field in the bus waveguide.

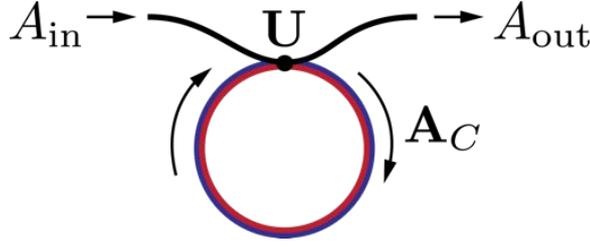

Figure S9: Multimode ring resonator formed by closing two ports into loops.

First, we assume that the power is conserved through the coupler, giving $\mathbf{A}_{\text{out}} = \mathbf{U} \cdot \mathbf{A}_{\text{in}}$. Secondly, we use the fact that the input fields in waveguide 1 and 2 are related to the output fields by free propagation, i.e. $A_{1,\text{in}} = e^{(ik_1 - \alpha_1/2)L} A_{1,\text{out}}$.

To compactly describe this physics we express the field as $\mathbf{A}_{\text{in}} = (A_{\text{in}}, \mathbf{A}_{C,\text{in}})$, where $\mathbf{A}_C = (A_2, A_1)$ are the two modes in the ring, we introduce a propagation matrix $\mathbf{T}_C$ for the modes in the ring cavity only i.e.,

$$\mathbf{T}_C(\ell) = \begin{pmatrix} e^{(ik_2 - \alpha_2/2)\ell} & 0 \\ 0 & e^{(ik_1 - \alpha_1/2)\ell} \end{pmatrix}, \tag{81}$$

and we express the scattering matrix as

$$\mathbf{U} = \begin{pmatrix} r & -i\boldsymbol{\mu}^{\mathrm{T}} \\ -i\boldsymbol{\mu} & \mathbf{R} \end{pmatrix}, \tag{82}$$

where $\boldsymbol{\mu}$ is the multi-mode generalization of the ring-bus coupling (*42*).

Using this notation we find that the relationship between the input and the output field in the bus waveguide is given by

$$\begin{pmatrix} A_{\text{out}} \\ \mathbf{A}_{C,\text{out}} \end{pmatrix} = \begin{pmatrix} r & -i\boldsymbol{\mu}^{\mathrm{T}} \\ -i\boldsymbol{\mu} & \mathbf{R} \end{pmatrix} \begin{pmatrix} A_{\text{in}} \\ \mathbf{A}_{C,\text{in}} \end{pmatrix}, \tag{83}$$

and the relationship between the cavity fields is

$$\mathbf{A}_{C,\text{in}} = \mathbf{T}_C(L) \cdot \mathbf{A}_{C,\text{out}}. \tag{84}$$

Solving these coupled equations we find the relationship between the input and the output fields in the bus, we find

$$A_{\text{out}} = \left[ r - \boldsymbol{\mu}^{\mathrm{T}} \cdot \mathbf{T}_C(L) \cdot (1 - \mathbf{R} \cdot \mathbf{T}_C(L))^{-1} \cdot \boldsymbol{\mu} \right] A_{\text{in}} \tag{85}$$

$$\mathbf{A}_{C,\text{out}} = -i(1 - \mathbf{R} \cdot \mathbf{T}_C(L))^{-1} \cdot \boldsymbol{\mu} A_{\text{in}}. \tag{86}$$

Once the scattering matrix for the six-port coupler, the group velocities, and the spatial decay rates are known, these results provide the power transmission spectrum given by

$$\frac{P_{\text{out}}}{P_{\text{in}}} = \left| r - \boldsymbol{\mu}^{\mathrm{T}} \cdot \mathbf{T}_C(L) \cdot (1 - \mathbf{R} \cdot \mathbf{T}_C(L))^{-1} \cdot \boldsymbol{\mu} \right|^2. \tag{87}$$

### 3.4 Relation between intracavity pump power and transmitted pump power

Using the relations, we can calculate the intracavity field strength, which is given by

$$|\mathbf{A}_{cav}|^2 = |\mathbf{r} - \boldsymbol{\mu}^{\mathrm{T}} \cdot (1 - \mathbf{R} \cdot \mathbf{T_C(L)})^{-1} \cdot \boldsymbol{\mu}|^2 \frac{1}{|\boldsymbol{\mu}^{\mathrm{T}} \cdot \boldsymbol{\mu}|} |(1 - \mathbf{R} \cdot \mathbf{T_C(L)}) \cdot \mathbf{A}_{out}|^2. \qquad (88)$$

Using this relation the transmitted pump power can be used to determine the amplitude of the pump field inside the resonator which is a key parameter for our numerical modeling.

### 3.5 Relation between intracavity laser power to output laser power

In order to fully understand Brillouin lasing in a ring resonator we establish the connection between the intracavity laser power and the output laser power coupled to the bus waveguide. This connection can be made using the ring resonator theory given above.

When operating as a Brillouin laser there is no Stokes light supplied to the resonator via the bus waveguide. Rather, pump light at a different frequency is injected into the resonator, and through Brillouin scattering leads to the generation of Stokes light in the resonator at a different frequency. By setting $A_{in}$ to zero we find the

$$A_{out,laser} = -i\boldsymbol{\mu}^{\mathrm{T}} \cdot \mathbf{A}_{C,in,laser}. \qquad (89)$$

Here, it is important to stress a crucial point. Ideally, our Brillouin laser operates by scattering injected pump light from the first excited mode in the ring to the fundamental mode, and the light in these two fields, of distinct frequencies, is confined to each of these respective modes. Hence, one may conclude that relation between the out coupled laser power and the intracavity power is simply given by $A_{out} = -i\boldsymbol{\mu} \cdot (0, A_1)$. However, unitary scattering through a three port coupler necessarily couples the first excited and the fundamental modes, and therefore Stokes light will be distributed in both modes. Therefore, the accurate connection between the measured laser power and output power necessarily contains the coherent addition of the amplitudes of the Stokes fields in both modes as given above.

### 3.6 A minimal model of the six-port coupler scattering matrix

In this section we discuss the scattering matrix for the six-port coupler. The most general $3 \times 3$ unitary and reciprocal scattering matrix is specified by 6 degrees of freedom. Given such a large parameter space, it is ideal to determine the scattering matrix through independent transmission measurements, or by finite difference time-domain (FDTD) simulations. However, these two approaches offer challenges in many instances. It is often not possible to directly measure the scattering matrix in on-chip systems (such as ours), and FDTD is limited by non-idealities in real systems such as sidewall roughness.

To circumvent these challenges we adopt a simple model of our six-port coupler. This model is constructed from three sequential unitary scattering matrices (i.e. $\mathbf{U} = \mathbf{U}_1 \cdot \mathbf{U}_2 \cdot \mathbf{U}_1$), the first coupling light between the bus and mode 1 of the ring $\mathbf{U}_1$, the second scattering the bus to mode 2 $\mathbf{U}_2$, and a third scattering the bus to mode 1 $\mathbf{U}_1$ (Fig. S10). Unitarity of each matrix conserves power through the

26

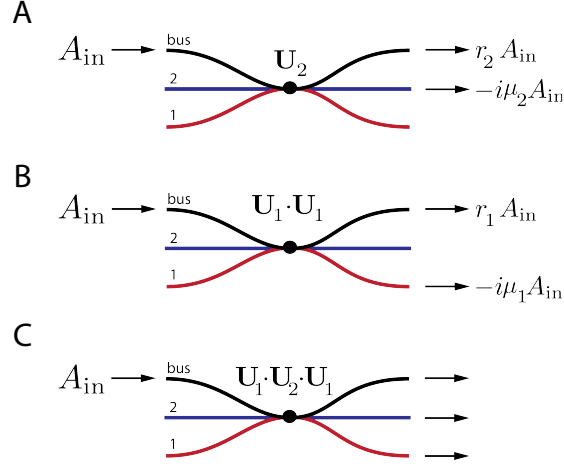

Figure S10: Minimal model of general reciprocal six-port scattering matrix. A subset of such scattering matrices are described by three sequential unitary transformations: (A) scattering between the bus and port 2, and (B) scattering between the bus and port 1. (C) Resultant six-port scattering from scattering in (B), (A), and (B) sequentially.

coupler and symmetry ensures reciprocity. These two scattering matrices are given by

$$\mathbf{U}_2 = \begin{pmatrix} r_2 & -i\mu_2 & 0 \\ -i\mu_2 & r_2 & 0 \\ 0 & 0 & 1 \end{pmatrix} \tag{90}$$

$$\mathbf{U}_1 = \begin{pmatrix} \sqrt{\frac{1+r_1}{2}} & 0 & -i\frac{\mu_1}{\sqrt{2(1+r_1)}} \\ 0 & 1 & 0 \\ -i\frac{\mu_1}{\sqrt{2(1+r_1)}} & 0 & \sqrt{\frac{1+r_1}{2}} \end{pmatrix} \tag{91}$$

$$\mathbf{U}_1 \cdot \mathbf{U}_1 = \begin{pmatrix} r_1 & 0 & -i\mu_1 \\ 0 & 1 & 0 \\ -i\mu_1 & 0 & r_1 \end{pmatrix}, \tag{92}$$

where unitarity requires that $\mu_1 = \sqrt{1-r_1^2}$ and $\mu_2 = \sqrt{1-r_2^2}$, and where we have factored the "rotation by $\mu_1$" from the bus to mode 1 into two.

While this approach fails to capture the most general six-port scattering physics, it does capture the salient features of six-port couplers and offers a simple tractable model described entirely by two free parameters.

Before moving on, we mention that we have explored the most general model to describe a lossless reciprocal 6-port coupler. However, the large number of parameters (5 + global phase) are underconstrained by our transmission spectra measurements, with families of parameters giving equally valid descriptions of our data. Furthermore, we find that the output of this general model yield properties of the optical modes in the ring in quantitative agreement with the simple model given above.

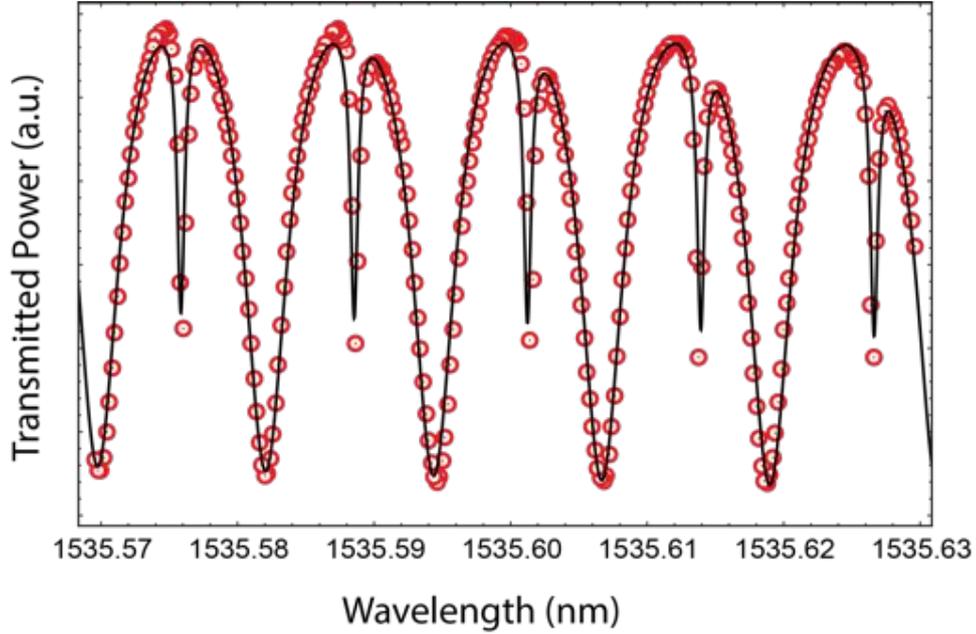

Figure S11: Transmitted pump power as a function of wavelength (open red circles). The black line is a fit of our transmission spectrum model to these data.

## 3.7 Inter-modal ring resonator characterization

Having established the basic physics of multi-mode ring resonators, we analyze transmission spectra measurements of our system to obtain the group velocities, frequencies, and loaded optical decay rates (see section 6.1). We model the transmission spectra using Eq. 87 and using the minimal mode for a three port coupler described by the scattering matrix given in Eq. 90. In addition, a slow wavelength dependent shift in the average transmitted power is added to our model to account for changes in the coupling efficiency as a function of wavelength (*19*). The result of fitting this model to the measured transmission spectra are shown in Fig. S11. This fitting procedure, in addition to SIMS measurements in linear waveguides, yields the model parameters describing our multimode Brillouin resonator given in section 6.1.

The decay rates presented in section 6.1 were obtained by an eigenfrequency analysis of the ring resonator, using the fitted scattering matrix. Eqs. 83 and 84 imply that the optical field in the ring is described by a set of lossy modes, characterized by complex eigenfrequencies. A subset of these eigenfrequencies are separated by 6.02 GHz, the frequency of the phonon that participates in Brillouin scattering. These pairs of modes are responsible for Brillouin lasing. Once these modes are identified, the decay rate can be obtained from the imaginary parts of the eigenfrequency.

The obtained decay rates given above can be checked by assuming negligible crosstalk between the optical modes in the ring, which is approximately valid in our system. When this is true the parameters $r_1$ and $r_2$ can be added to the bare decay rates, i.e. the decay rates of the fields in a linear waveguide, to estimate the loaded decay rate which accounts for coupling loss. These relations are given by

$$\Delta\nu_{\mathrm{p}} \approx \frac{v_{\mathrm{g,p}}}{2\pi}(\alpha_{\mathrm{p}} - \frac{2}{L}\ln r_2) = 489 \text{ MHz} \tag{93}$$

$$\Delta\nu_{\mathrm{s}} \approx \frac{v_{\mathrm{g,s}}}{2\pi}(\alpha_{\mathrm{s}} - \frac{2}{L}\ln r_1) = 86 \text{ MHz}, \tag{94}$$

showing the self-consistency of our analysis. The small difference between the tabulated decay rates and the decay rates given above is due to a small amout of coupling between the ring resonator modes.

## 3.8 Laser power conversion factor

Our scattering theory treatment of the ring resonator allows us to estimate the ratio of intracavity laser power to output laser power, the amount exiting the resonator on the bus waveguide. This parameter is necessary to compare our simulations of the laser physics to our measurements. For the modes responsible for lasing we find a power conversion factor of 0.03. However, this parameter is relatively unconstrained by our transmission spectra data as it depends strongly on the relative phase between the two optical spatial modes exiting the 6-port coupler (as discussed in section 3.5). By fitting various forms of the scattering matrix to the transmission spectrum, we have found a range of conversion factor values that are consistent with the measured transmission spectra. For this reason, we will treat this conversion factor as a fit parameter in our theory-experiment comparisons. We find that our measurements of slope efficiency and the pump-Stokes beat-note linewidth are best described for a power conversion factor of 0.0075.

# 4 Data analysis

In this section, we describe the methods used to analyze the experimental data.

We use high-resolution heterodyne spectroscopy to characterize the power and linewidth of the emitted Stokes light. This involves synthesizing an optical local oscillator (LO), which is 44 MHz blue-detuned from the incident pump field, and combining it with the emitted Stokes light on a high-speed photoreceiver (see Fig. 2A of Report).

Through careful calibration of the detector response and the optical local oscillator power, these microwave spectra provide an accurate measure of the Stokes power as well as the optical linewidth relative to that of the pump wave. By simultaneously measuring the power and the heterodyne beat note from a reference pump wave (beat note produced by the reference pump wave and the LO), we can calibrate the power spectral density of the microwave spectrum—thus allowing us to determine the emitted Stokes power. Also, because the optical LO is synthesized from the pump laser, the heterodyne beat note gives an accurate measure of the emitted Stokes linewidth when the emitted Stokes spectrum is broader than the pump laser (below threshold). Above threshold (when the beat note linewidth is much smaller than that of the pump linewidth), we use the sub-coherence self-heterodyne technique to characterize the pump-Stokes coherence, as detailed in section 2 of the supplement.

## 4.1 Slope efficiency

In Fig. 2B of the main text, we plot the Stokes power as a function of injected pump power. The output (on-chip) Stokes power is determined through a calibrated heterodyne measurement. While performing this heterodyne spectroscopy, we simultaneously measure the input and transmitted pump powers

through the chip. By monitoring the input and transmitted pump powers and using the racetrack resonator properties analyzed in Section 3 (including coupling and linear loss), we infer the detuning of the pump from resonance and estimate the intra-cavity pump powers. The data presented in Fig. 2B represent the measured Stokes powers (left $y$-axis) as a function of intracavity pump power (top $x$-axis). For comparison, the bottom $x$-axis assumes a zero-detuned resonant enhancement factor ($\approx 1.8$), showing the effective input pump power. The right $y$-axis indicates the estimated intracavity Stokes power assuming the intracavity to bus waveguide power conversion factor discussed in Section 3.8 (0.0075).

## 4.2 Linewidth measurements

This section discusses the analysis of the data presented in Fig. 3C of the Report.

### 4.2.1 Heterodyne measurements

Below threshold, we use standard heterodyne spectroscopy to determine the linewidth of the phonon field (see Fig. 3A,C of the Report). To avoid spectral distortion due to pump mode hopping, we acquire rapid spectral traces at different output Stokes powers. We then apply a moving average (with a bin size smaller than linewidth to avoid distortion) to spectrally smooth the data for consistent Lorentzian fits. The data presented in Fig. 3C are acquired using a resolution bandwidth of 20 kHz. Above threshold, the phonon linewidths are resolution-bandwidth limited. To simplify the representation of the data in Fig. 3C, we only include spectral measurements below threshold.

### 4.2.2 Sub-coherence self-heterodyne measurements

We determine the excess phase-noise linewidth of the Stokes wave relative to the pump wave (i.e., phonon linewidth) above threshold using the sub-coherence delayed self-heterodyne technique discussed in Section 2. To do this, we fit the theoretical spectrum (Eq. 73) to the measured spectra. We extract the phonon linewidth by fitting the power, pump linewidth, and phonon linewidth to the peak spectral density and the fringe-like portion of the spectrum (the feature highlighted in Fig. S7C), which is most sensitive to phonon phase noise. This method is only possible at large Stokes powers, where there is sufficient signal to noise so that the fringes do not become obscured by the noise floor.

As discussed in the Fig. S7 caption, instabilities in the fiber delay line of the interferometer add approximately $\sim 500$ Hz of phase noise to the sub-coherence self heterodyne measurements (obtained from a self-heterodyne measurement using an AOM as the frequency shifter). Therefore, to get an accurate measurement the phonon phase-noise within the silicon Brillouin laser, we normalize our sub-coherence self-heterodyne measured linewidths by subtracting the residual phase noise.

For a comparison with the spectral measurements below threshold (not resolution bandwidth limited), we convert the beat-note power into a peak spectral density ($x$-axis of Fig. 3C) by using the fact that the integral of a Lorentzian is proportional to the peak height multiplied by the width. In this way, we convert the sub-coherence spectrum into an equivalent Lorentzian that has a FWHM given by the measured linewidth and a peak height given by the peak spectral density. Thus, by dividing the Stokes power by the measured linewidth (and multiplying by $2/\pi$), we obtain the equivalent peak spectral density used for this comparison (plotted in Fig. 3C of the main text).

As discussed in Section 2.2, these sub-coherence self-heterodyne measurements demonstrate the pump-Stokes coherence produced by optical self-oscillation in this system. While these data permit a measurement of the phonon linewidth and corroborate the regime of dynamics (presence of fringes and measurements of phonon linewidths below that of the pump, see Section 2.2.a-b), the data do not yield sufficient precision to clearly determine the linewidth narrowing trend as function of total Stokes power. To attain the necessary level of precision, this type of analysis would likely require more sophisticated

control of all variables that influence laser oscillation. Nevertheless, these data establish an upper bound on the phonon linewidths above threshold and clearly reveal the pump-Stokes phase coherence unique to this limit of Brillouin laser oscillation.

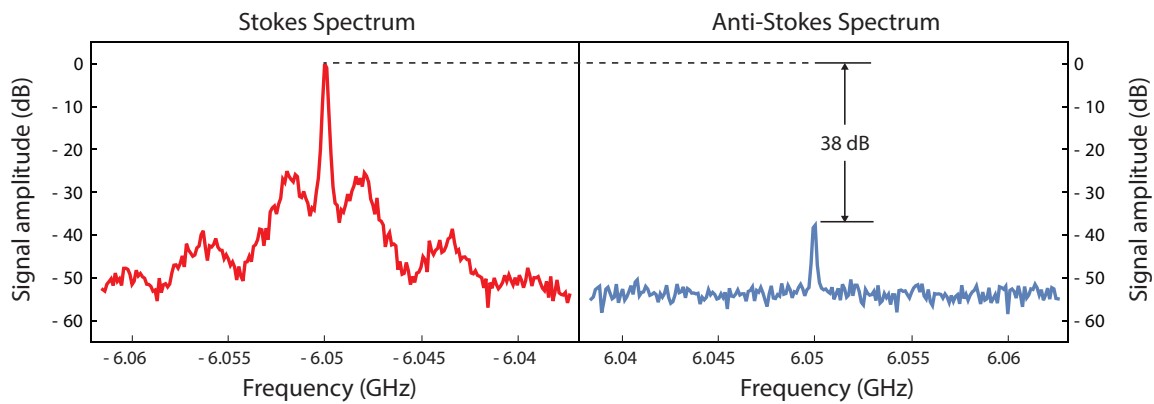

Figure S12: Above threshold Stokes and anti-Stokes spectra demonstrating 38 dB of anti-Stokes suppression.

# 5    Additional measurements and considerations

## 5.1    Anti-Stokes suppression

The inter-modal Brillouin scattering process provides intrinsic Stokes-anti-Stokes symmetry breaking and single-sideband amplification. This unique property of the SIMS gain mechanism permits a large degree of anti-Stokes suppression (38 dB) of the output spectrum. As discussed in the main text, the Stokes phonons (generated from the pump and Stokes waves propagating in distinct optical spatial modes) cannot mediate the anti-Stokes process. Above threshold, the strong and mutually coherent pump and Stokes waves set up an acoustic grating, which—if the pump propagates purely in the anti-symmetric mode—phase matches only to the Stokes process. However, in practice a small fraction of the pump light couples into the wrong (symmetric) spatial mode; pump light propagating in this spatial mode can interact with the strong acoustic grating and phase match in an anti-Stokes process, thus resulting in a small anti-Stokes sideband.

## 5.2    Output coupling and slope efficiency

Increasing the output Stokes coupling can enhance the slope efficiency, but this will also degrade the optical Q-factor and consequently raise the threshold pump power. However, because this Brillouin laser is under-coupled for the Stokes wave, the dominant source of loss limiting the Q-factor is the intrinsic propagation loss of the symmetric mode. In this Brillouin laser, therefore, increasing the output coupling can dramatically enhance the slope efficiency while only marginally raising the threshold power. Using Eq. 44, we estimate that increasing the output coupling to 16% could increase the slope efficiency by 10 fold. To illustrate this dependence, Fig. S13 shows a comparison between the slope efficiencies with various output Stokes couplings.

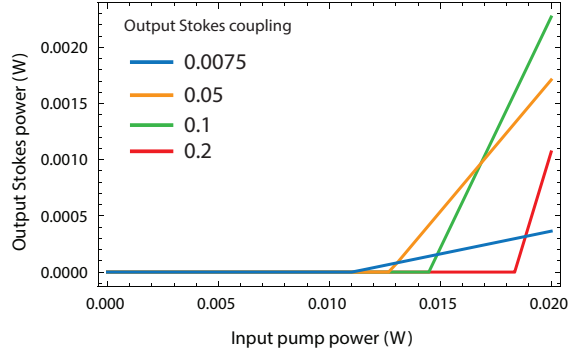

Figure S13: Theoretical slope efficiency (Eq. 44) with various output Stokes couplings. The blue trace (output coupling = 0.0075) represents the predicted slope efficiency for the laser studied in this work.

## 5.3 Additional sub-coherence self-heterodyne measurements with open-loop laser

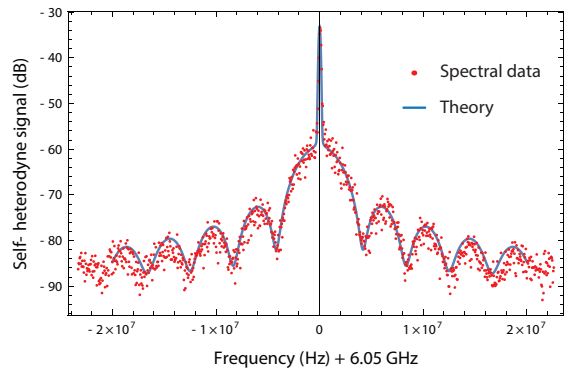

Figure S14: Typical self-heterodyne spectrum and a fitted theoretical spectrum (see Section 2.2.1) of Brillouin laser using an open-loop pump laser (Pure Photonics PPCL300) . This theoretical fit corresponds to a pump laser linewidth of 32 kHz and an excess phase noise linewidth (phonon linewidth) of 1.73 kHz, demonstrating (1) the pump-Stokes phase coherence unique to this limit of Brillouin dynamics and (2) a phonon linewidth far below that of the pump laser.

We found it most practical to use the Agilent 81600B as the pump laser for our measurements due to its frequency agility and wavelength sweeping features, which allow us to quickly and systematically characterize the system. It does, however, exhibit additional phase-noise characteristics (slight deviations from theoretical sub-coherence spectrum, see Fig. S7); these features are due to low-frequency locking electronics and are common in lasers that use active locking for stabilization. To crosscheck these measurements, we have also used a thermally-tunable, open-loop diode laser that exhibits superb agreement with sub-coherence spectrum. Using this laser, we observe exceptional agreement between the measured and predicted spectrum (see Section 2.2.b), corroborating the results obtained from our initial analysis.

In Fig. S14, we have included an example sub-coherence self-heterodyne spectrum of our silicon

Brillouin laser using an open-loop pump laser (Pure Photonics PPCL300). This particular spectrum was measured using a laser device with a slightly different length (3.06 cm circumference). Note that sidebands due to frequency stabilization electronics are absent from this measurement. The fitted spectrum in Fig. S14 corresponds to a phonon linewidth is 1.73 kHz (and pump linewidth of 32 kHz), revealing similar phonon linewidths under similar conditions using the Agilent 81600B.

## 5.4   Inter-modal Brillouin laser concept

This Brillouin laser design—made possible by the forward inter-modal scattering process—is quite different from the geometries of previous Brillouin laser systems. These fundamental distinctions in device concept alleviate a number of challenges facing on-chip integration of Brillouin lasers. Since conventional Brillouin lasers are based on backward SBS, they invariably produce Stokes laser emission in the backward direction. In many cases, this poses a significant challenge for integration; such Brillouin lasers would likely require on-chip circulators and isolators to protect the pump laser from unwanted feedback. However, the absence of mature isolator and circulator technology remains a serious obstacle facing the field of integrated photonics. By contrast, this silicon Brillouin laser uses a forward Brillouin scattering process, which produces Stokes emission in the forward direction. In this way, the inter-modal forward Brillouin laser geometry avoids the need for isolators and circulators. Moreover, since the forward Brillouin scattering process is created using structural control, this device concept is quite general and in principle, can be translated to a number of material systems.

Beyond its suitability for integration, this inter-modal Brillouin laser yields flexibility in design and control of dynamics in ways that are difficult to achieve in Brillouin lasers that use backward stimulated Brillouin scattering. In particular, the multimode nature of the inter-modal Brillouin process permits independent tuning of the dissipation rates and cavity enhancements for the two spatial modes, allowing us to precisely shape the energy transfer between pump and Stokes waves. This is different from conventional Brillouin lasers, where the pump and Stokes light propagate in the same spatial mode and have nearly identical dissipation rates. If, through Brillouin laser oscillation, the Stokes wave becomes strong enough in conventional systems, it can reach the lasing threshold to produce cascaded energy transfer to successive Stokes orders. In contrast, this inter-modal system allows us to precisely control the degree to which cascading can or cannot occur. For example, by tailoring the Q-factor of the pump wave, we can precisely engineer the Stokes-power threshold required for cascaded energy transfer. Alternatively, we can completely suppress cascading by engineering the mode spectrum such that the 2$^{\text{nd}}$ Stokes order wave is not supported by a cavity mode. This is the strategy used in our current system to suppress cascading.

In addition, the tailorability afforded by the inter-modal Brillouin scattering process allows us to optimize this laser design according to any number of constraints. In contrast with Raman and Kerr non-linearities, whose properties are principally governed by the waveguide material, Brillouin nonlinearities are exceptionally tailorable through device geometry. In fact, it is possible to engineer a range of essential characteristics of this Brillouin nonlinearity, including resonance frequency, acoustic dissipation rate, and Brillouin gain (*15, 16*). Beyond modifying the waveguide properties, the multi-mode nature of this inter-modal Brillouin interaction permits great flexibility in the design of the laser cavity. This is because this inter-modal Brillouin laser topology permits phase-matching to be satisfied over a tremendous range of resonator dimensions. By reducing the footprint and increasing the cavity finesse, we can significantly increase the resonant pump-power enhancement, thereby dramatically lowering the laser threshold. For example, a $100\times$ size reduction could permit resonant power enhancements of $\sim 180$, pushing the laser threshold below 100 $\mu$W.

33

# 6 Methods

## 6.1 Experimental Parameters

| Optical properties | value |
|---|---|
| L | 4.576 cm |
| FSR (symmetric) | 1.614 GHz |
| FSR (anti-symmetric) | 1.570 GHz |
| $r_2$ | 0.54 |
| $\mu_2$ | 0.84 |
| $r_1$ | 0.97 |
| $\mu_1$ | 0.24 |
| $\alpha_2$ | 15.8 m$^{-1}$ |
| $\alpha_1$ | 6.0 m$^{-1}$ |
| $v_{g,2}$ | 7.163 $\times 10^7$ m/s |
| $v_{g,1}$ | 7.385 $\times 10^7$ m/s |
| $\gamma_2$ | $2\pi$ 481 MHz |
| $\gamma_1$ | $2\pi$ 83 MHz |
| I$_{\text{threshold}}$ | $5.4 \times 10^6$ Wcm$^{-2}$ |
| Pump linewidth | 13 kHz |

Table S1: Optical parameters

| Acoustic properties | value |
|---|---|
| $\Omega_{\text{b}}$ | $2\pi$ 6.02 GHz |
| $g$ | 11.1 kHz |
| $\Gamma_0$ | $2\pi$ 13.1 MHz |
| $q$ | $4.5 \times 10^5$ m$^{-1}$ |
| $v_{\text{b,group}}$ | 826 m/s |
| $v_{\text{b,phase}}$ | $8.4 \times 10^4$ m/s |

Table S2: Acoustic parameters

| Nonlinear optical properties | value |
|---|---|
| Brillouin Gain<br>$G_b$ | $400\ \mathrm{W}^{-1}\mathrm{m}^{-1}$ |
| Kerr coefficients<br>$\gamma_k^{11}$<br>$\gamma_k^{22}$<br>$\gamma_k^{12} = \gamma_k^{21}$ | <br>$74 \pm 11\ \mathrm{m}^{-1}\mathrm{W}^{-1}$<br>$71 \pm 11\ \mathrm{m}^{-1}\mathrm{W}^{-1}$<br>$44 \pm 7\ \mathrm{m}^{-1}\mathrm{W}^{-1}$ |
| TPA coefficients<br>$\beta^{11}$<br>$\beta^{22}$<br>$\beta^{12} = \beta^{21}$ | <br>$34 \pm 10\ \mathrm{m}^{-1}\mathrm{W}^{-1}$<br>$30 \pm 9\ \mathrm{m}^{-1}\mathrm{W}^{-1}$<br>$20 \pm 6\ \mathrm{m}^{-1}\mathrm{W}^{-1}$ |
| FCA coefficients<br>$\gamma^{111}$<br>$\gamma^{222}$<br>$\gamma^{211} \approx \gamma^{122}$ | <br>$900 \pm 400\ \mathrm{m}^{-1}\mathrm{W}^{-2}$<br>$720 \pm 430\ \mathrm{m}^{-1}\mathrm{W}^{-2}$<br>$310 \pm 200\ \mathrm{m}^{-1}\mathrm{W}^{-2}$ |

Table S3: Nonlinear optical parameters

## 6.2 Device fabrication

The Brillouin laser devices were fabricated on a silicon-on-insulator chip with a 3 μm oxide layer beneath a 215 nm-thick crystalline silicon layer. The waveguides were patterned using electron beam lithography on hydrogen silsesquioxane photoresist. After development in Microposit™ MF™-312 developer, a $Cl_2$ reactive ion etch (RIE) was used to create the ridge structures. Following a solvent cleaning step, slot structures were written with electron beam lithography on ZEP520A photoresist and $Cl_2$ RIE. The device was then wet released via a 49% hydrofluoric acid etch of the oxide layer. The laser structure reported here consists of two regions of 436 suspended segments each and has a total length of 4.576 cm.

# References

1. G. T. Reed, G. Mashanovich, F. Y. Gardes, D. J. Thomson, Silicon optical modulators. *Nat. Photonics* **4**, 518 (2010).

2. D. Liang, J. E. Bowers, Recent progress in lasers on silicon. *Nat. Photonics* **4**, 511 (2010).

3. J. Leuthold, C. Koos, W. Freude, Nonlinear silicon photonics. *Nat. Photonics* **4**, 535 (2010).

4. A. Alduino, M. Paniccia, Interconnects: Wiring electronics with light. *Nat. Photonics* **1**, 153 (2007).

5. K. D. Vos, I. Bartolozzi, E. Schacht, P. Bienstman, R. Baets, Silicon-on-insulator microring resonator for sensitive and label-free biosensing. *Opt. Express* **15**, 7610 (2007).

6. O. Boyraz, B. Jalali, Demonstration of a silicon Raman laser. *Opt. Express* **12**, 5269 (2004).

7. H. Rong, *et al.*, A continuous-wave Raman silicon laser. *Nature* **433**, 725 (2005).

8. H. Takuma, D. Jennings, Stimulated Brillouin scattering in the off-axis resonator. *Appl. Phys. Lett.* **5**, 239 (1964).

9. V. Lecoeuche, D. J. Webb, C. N. Pannell, D. A. Jackson, Brillouin based distributed fibre sensor incorporating a mode-locked Brillouin fibre ring laser. *Opt. Commun.* **152**, 263 (1998).

10. S. P. Smith, F. Zarinetchi, S. Ezekiel, Narrow-linewidth stimulated Brillouin fiber laser and applications. *Opt. Lett.* **16**, 393 (1991).

11. W. Loh, *et al.*, Dual-microcavity narrow-linewidth Brillouin laser. *Optica* **2**, 225 (2015).

12. I. V. Kabakova, *et al.*, Narrow linewidth Brillouin laser based on chalcogenide photonic chip. *Opt. Lett.* **38**, 3208 (2013).

13. J. Li, M.-G. Suh, K. Vahala, Microresonator Brillouin gyroscope. *Optica* **4**, 346 (2017).

14. P. T. Rakich, C. Reinke, R. Camacho, P. Davids, Z. Wang, Giant enhancement of stimulated Brillouin scattering in the subwavelength limit. *Phys. Rev. X* **2**, 011008 (2012).

15. H. Shin, *et al.*, Tailorable stimulated Brillouin scattering in nanoscale silicon waveguides. *Nat. Commun.* **4**, 1944 (2013).

16. R. Van Laer, B. Kuyken, D. Van Thourhout, R. Baets, Interaction between light and highly confined hypersound in a silicon photonic nanowire. *Nat. Photonics* **9**, 199 (2015).

17. E. A. Kittlaus, H. Shin, P. T. Rakich, Large Brillouin amplification in silicon. *Nat. Photonics* **10**, 463 (2016).

18. R. V. Laer, A. Bazin, B. Kuyken, R. Baets, D. V. Thourhout, Net on-chip Brillouin gain based on suspended silicon nanowires. *New J. Phys.* **17**, 115005 (2015).

19. E. A. Kittlaus, N. T. Otterstrom, P. T. Rakich, On-chip inter-modal Brillouin scattering. *Nat. Commun.* **8**, 15819 (2017).

20. Y. A. Espinel, F. G. Santos, G. O. Luiz, T. M. Alegre, G. S. Wiederhecker, Brillouin optomechanics in coupled silicon microcavities. *Sci. Rep.* **7**, 43423 (2017).

21. M. S. Kang, A. Brenn, P. St.J. Russell, All-optical control of gigahertz acoustic resonances by forward stimulated interpolarization scattering in a photonic crystal fiber. *Phys. Rev. Lett.* **105**, 153901 (2010).

22. G. Bahl, J. Zehnpfennig, M. Tomes, T. Carmon, Stimulated optomechanical excitation of surface acoustic waves in a microdevice. *Nat. Commun.* **2**, 403 (2011).

23. Materials, methods are available as supplementary materials .

24. J. Li, H. Lee, T. Chen, K. J. Vahala, Characterization of a high coherence, Brillouin microcavity laser on silicon. *Opt. Express* **20**, 20170 (2012).

25. A. Debut, S. Randoux, J. Zemmouri, Linewidth narrowing in Brillouin lasers: Theoretical analysis. *Phys. Rev. A* **62**, 023803 (2000).

26. I. S. Grudinin, H. Lee, O. Painter, K. J. Vahala, Phonon laser action in a tunable two-level system. *Phys. Rev. Lett.* **104**, 083901 (2010).

27. W. C. Jiang, X. Lu, J. Zhang, Q. Lin, High-frequency silicon optomechanical oscillator with an ultralow threshold. *Opt. Express* **20**, 15991 (2012).

28. K. Hill, B. Kawasaki, D. Johnson, CW Brillouin laser. *Appl. Phys. Lett.* **28**, 608 (1976).

29. K. S. Abedin, Single-frequency Brillouin lasing using single-mode As2Se3 chalcogenide fiber. *Opt. Express* **14**, 4037 (2006).

30. B. Morrison, *et al.*, Compact Brillouin devices through hybrid integration on silicon. *Optica* **4**, 847 (2017).

31. R. W. Boyd, *Nonlinear optics* (Academic press, 2003).

32. G. P. Agrawal, *Nonlinear fiber optics* (Academic press, 2007).

33. J. Sipe, M. Steel, A Hamiltonian treatment of stimulated Brillouin scattering in nanoscale integrated waveguides. *New J. Phys* **18**, 045004 (2016).

34. P. Kharel, R. Behunin, W. Renninger, P. Rakich, Noise and dynamics in forward Brillouin interactions. *Phys. Rev. A* **93**, 063806 (2016).

35. R. Bonifacio, L. Lugiato, Optical bistability and cooperative effects in resonance fluorescence. *Phys. Rev. A* **18**, 1129 (1978).

36. A. Yariv, *Quantum electronics, 3rd* (John Wiley & Sons, 1989).

37. K. J. Vahala, Back-action limit of linewidth in an optomechanical oscillator. *Phys. Rev. A* **78**, 023832 (2008).

38. L. Ge, Y. D. Chong, A. D. Stone, Steady-state ab initio laser theory: generalizations and analytic results. *Physical Review A* **82**, 063824 (2010).

39. W. Renninger, P. Kharel, R. Behunin, P. Rakich, Bulk crystalline optomechanics. *Nat. Phys.* **pp** (2018).

40. L. Richter, H. Mandelberg, M. Kruger, P. McGrath, Linewidth determination from self-heterodyne measurements with subcoherence delay times. *IEEE J. Quantum Electron.* **22**, 2070 (1986).

41. M. Van Exter, S. Kuppens, J. Woerdman, Excess phase noise in self-heterodyne detection. *IEEE J. Quantum Electron.* **28**, 580 (1992).

42. B. E. Little, S. T. Chu, H. A. Haus, J. Foresi, J. Laine, Microring resonator channel dropping filters. *J. Lightwave Technol.* **15**, 998 (1997).



\end{document}